\documentclass[11pt,a4paper]{article}
\pdfoutput=1
\usepackage{jheppub}
\usepackage[...]{youngtab}
\usepackage{appendix}
\def\Bbb{\mathbb}
\usepackage{amsfonts}
\usepackage[usenames,dvipsnames]{xcolor}
\usepackage{color}
\definecolor{Red}{rgb}{1,0,0}
\def\BZ{\mbox{$\Bbb Z$}}

\def\Bb{{\cal B}_2}
\def\Bc{{\cal B}_3}
\def\Bd{{\cal B}_4}

\catcode`@=11 \@addtoreset{equation}{section} \catcode`@=12

\newcommand{\sid}{\begin{equation}}
\newcommand{\sidd}{\end{equation}}
\usepackage{pgf}
\usepackage{tikz}
\usepackage[utf8]{inputenc}
\usetikzlibrary{arrows,automata}
\usetikzlibrary{positioning}
\tikzset{state/.style={rectangle, rounded corners, draw=black, very thick, minimum height=2em, inner sep=2pt, text centered,},}
\title{Is toric duality a Seiberg-like duality in (2+1)-d ?}
\author{Siddharth Dwivedi}
\author{and P. Ramadevi}
\affiliation{Department of Physics, Indian Institute of Technology Bombay,\\
Mumbai 400 076, India}
\emailAdd{siddharth@phy.iitb.ac.in}
\emailAdd{ramadevi@phy.iitb.ac.in}
\abstract{We show that not all $(2+1)$ dimensional toric phases 
are Seiberg-like duals. Particularly, we work out superconformal indices 
for the toric phases of Fanos ${\cal{C}}_3$, ${\cal{C}}_5$ and ${\cal{B}}_2$.
We find that the indices for the two toric phases of Fano ${\cal{B}}_2$ do not match, which implies that they are not Seiberg-like duals. We also take the route of  acting Seiberg-like duality transformation on toric quiver Chern-Simons theories to obtain dual quivers. We
study two examples and show that  Seiberg-like dual quivers are not always toric quivers.}
\keywords{} 
\arxivnumber{1401.2767}
\begin{document}
\maketitle
\flushbottom
\section{Introduction}
Works on AdS/CFT has shown that $D3$ ($M2$)-branes probing toric Calabi-Yau threefold (fourfold) give rise to ${\cal{N}}=1$ (${\cal{N}}=2$) toric quiver gauge theories on the worldvolume of the branes. A toric quiver gauge theory has a superpotential ($W$) in which all the matter fields appear exactly twice, once with positive and once with a negative term respectively. We will restrict to only those toric quivers in which the gauge group associated to each node is same (see footnote 5 of ref. \cite{Franco:2005sm}). There exists an algorithm for finding the toric data of Calabi-Yau threefold from the ${\cal{N}}=1$ quiver gauge theory, which is known in the literature as \emph{forward} algorithm \cite{Feng:2000mi}. This algorithm can be extended in the context of $M2$-branes to obtain the Calabi-Yau fourfold toric data from the ($2+1$) dimensional quiver supersymmetric Chern-Simons theories with Chern-Simons levels \cite{Ueda:2008hx, Hanany:2008fj,Franco:2008um,Hanany:2008gx,Davey:2009et}. Conversely, we could obtain the matter content and superpotential of the quiver theories from the toric data using \emph{inverse} algorithm \cite{Feng:2000mi}. From inverse algorithm, we can find more than one quiver gauge theory sharing the same toric data. These quiver theories are called toric duals or phases. Several examples of toric phases both in (3+1) and (2+1) dimensions are available in the literature \cite{Feng:2001xr,Feng:2002zw,Davey:2009sr,Davey:2009qx}. All these quiver theories which were studied, admit the dimer tiling description \cite{Hanany:2005ve,Franco:2005rj,Hanany:2005ss,Kennaway:2007tq,Agarwal:2008yb,Hanany:2008cd}. From the inverse algorithm approach, we obtained toric phases \cite{Dwivedi:2011zm} which do not admit tiling.

\begin{figure}
	\centering
		\includegraphics[width=0.45\textwidth]{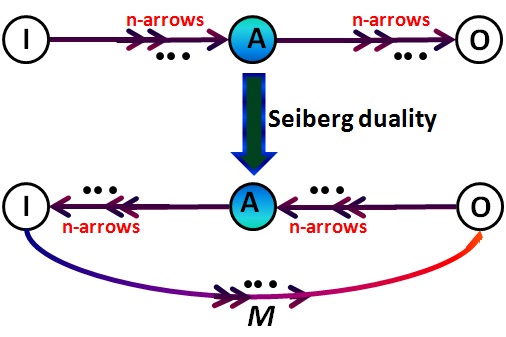}
	\caption{Seiberg duality transformation on node-A of the quiver diagram. Note that the incoming and outgoing arrows at node-A are reversed in the dual quiver. There are some mesonic fields $M$, although some of the fields can be massive due to the superpotential and has to be integrated out \cite{Feng:2001bn}.}
	\label{N=1-Seiberg-Duality}
\end{figure}

One of the main puzzle is to check whether the toric dual quiver theories in (3+1)-d or (2+1)-d are actually Seiberg duals \cite{Seiberg:1994pq,Berenstein:2002fi}. The ${\cal{N}}=1$ (3+1)-d quiver gauge theories which are toric duals always have same number of nodes in their quiver diagrams. If we apply the Seiberg duality transformation on a (3+1)-d quiver gauge theory, we can in principle, get a Seiberg dual theory as shown in figure \ref{N=1-Seiberg-Duality} and the new superpotential can be obtained following ref. \cite{Feng:2001bn}.
Applying the Seiberg duality on node-A, the corresponding gauge group of node-A will change from $SU(N)$ to $SU(nN-N)$, where $n$ is the number of incoming and outgoing arrows. For the toric quivers which we consider, the rank of node-A will remain $N$ if and only if $n=2$. So, we apply the Seiberg duality transformation on only those nodes in the quiver which have exactly two incoming and two outgoing arrows. In the earlier works of \cite{Feng:2001bn}, it was found that upon applying the Seiberg duality rules, the (3+1)-d toric phases transform into one another. Based on such results, it was conjectured \cite{Feng:2001bn,Beasley:2001zp} that ${\cal{N}}=1$ toric dual theories are also Seiberg duals.
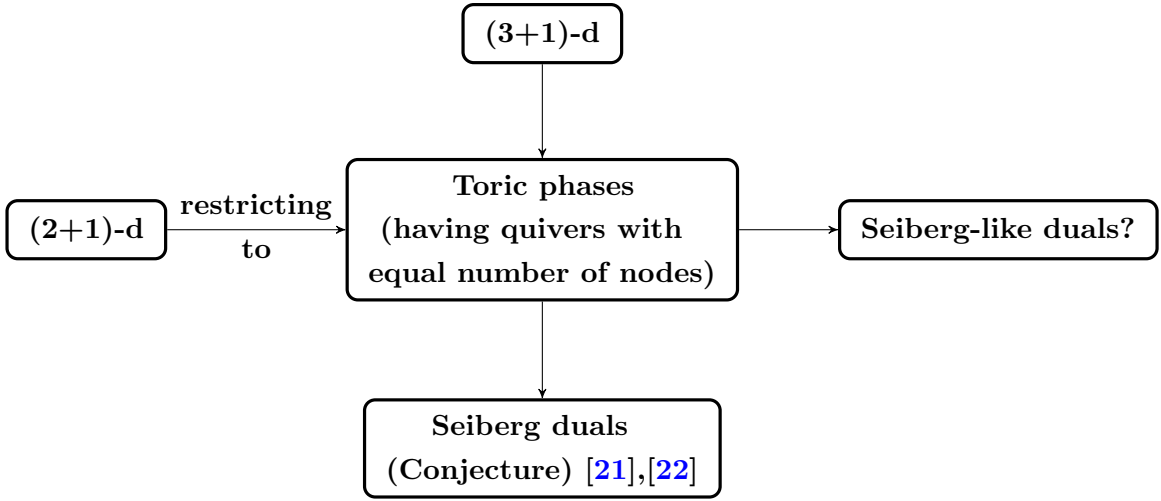
\begin{figure}
\begin{raggedleft}
\begin{tikzpicture}[->,>=stealth']
\node[state] (3D)
 {\begin{tabular}{l}
\textbf{(2+1)-d}
 \end{tabular}};
\node[state,    	
  right of=3D, 	
  node distance=6.0cm, 
  anchor=center] (TDUALS)
 {
\begin{tabular}{l}
\hspace*{1.1cm}\textbf{Toric phases} \\ \hspace*{0.15cm}\textbf{(having quivers with} \\ \textbf{equal number of nodes)}
\end{tabular}
};
\node[state,
  right of=TDUALS,
	node distance=6.0cm,
  anchor=center] (SEIBERG) 
 {
 \begin{tabular}{l}
  \textbf{Seiberg-like duals?} 
 \end{tabular}
 };
\node[state,
  above of=TDUALS,
  yshift= 1.6cm,
  anchor=center] (4D)
	{\begin{tabular}{l} 
\textbf{(3+1)-d}
\end{tabular}
 };
\node[state,
  below of=TDUALS,
	node distance=2.9cm,
  anchor=center] (SEIBERG?) 
{
 \begin{tabular}{l}
  \hspace*{0.6cm}\textbf{Seiberg duals} \\ \textbf{(Conjecture) \cite{Feng:2001bn},\cite{Beasley:2001zp}} 
 \end{tabular}
 };
 \path 
 (4D) edge (TDUALS)
 (TDUALS) edge (SEIBERG)
(TDUALS) edge (SEIBERG?)
 ;
\path  (3D) edge node[above]{\textbf{restricting}} node[below]{\textbf{to}} (TDUALS);

 
\end{tikzpicture}
\end{raggedleft}
\caption{In (3+1)-d, the toric duality implies Seiberg duality and does it hold in (2+1)-d?}
\label{flowchart}
\end{figure}

For (2+1) dimensional ${\cal{N}}=2$ quiver Chern-Simons theories, the situation is different. More details about the Seiberg-like duality in (2+1)-d can be found in \cite{RoblesLlana:2004nq,Benini:2011mf,Closset:2012eq,Agarwal:2012wd,Park:2013wta,Amariti:2011uw,Amariti:2009rb}. The main theme of this paper is to check whether the (2+1)-d toric phases, with equal number of nodes in their quiver diagram, are always Seiberg-like duals (see flowchart in figure \ref{flowchart}). Each node in a quiver Chern-Simons theory has an additional assignment of Chern-Simons level. The Seiberg-like duality transformation will still have same effect on the quiver diagram as shown in figure \ref{N=1-Seiberg-Duality} with the matter content and the superpotential transforming as in (3+1)-d case, but the ranks and the Chern-Simons levels for nodes I, O and A will change as given below \cite{Amariti:2009rb}: 
\begin{align}
U(N)_{k_I} & \longrightarrow U(N)_{k_I+k_A}~, \notag \\ 
U(N)_{k_A} & \longrightarrow U(nN-N+|k_A|)_{-k_A}~, \notag \\ 
U(N)_{k_O} & \longrightarrow U(N)_{k_O+k_A}~.
\label{S-like-rules}
\end{align}
So, we must apply the Seiberg-like duality on only those nodes for which $n=2$ and $k=0$ to study toric duals. For (2+1)-d quiver theories which have (3+1)-d parents \cite{Franco:2008um}, the Seiberg-like dual quiver diagram will be same as in (3+1)-d. Chern-Simons levels can be then appropriately assigned to each node consistent with the rules (\ref{S-like-rules}) so that the two Seiberg-like dual theories also share the same toric data \cite{Amariti:2009rb}. 

Note that for non-chiral or vector-like theories in which number of incoming and number of outgoing arrows between any two pair of nodes is same, Seiberg-like duality can be seen from the brane picture \cite{Amariti:2009rb}. For chiral theories, which have at least one pair of nodes (say nodes $i$ and $j$) such that number of arrows from $i \rightarrow j$ is different from number of arrows from $j \rightarrow i$, rules (\ref{S-like-rules}) can not be deduced from brane setup. However we can still apply (\ref{S-like-rules}) for Seiberg-like duality if we dualize only those nodes which have Chern-Simons level $k=0$, as mentioned in \cite{Amariti:2009rb}. 

\begin{figure}
	\centering
		\includegraphics[width=0.90\textwidth]{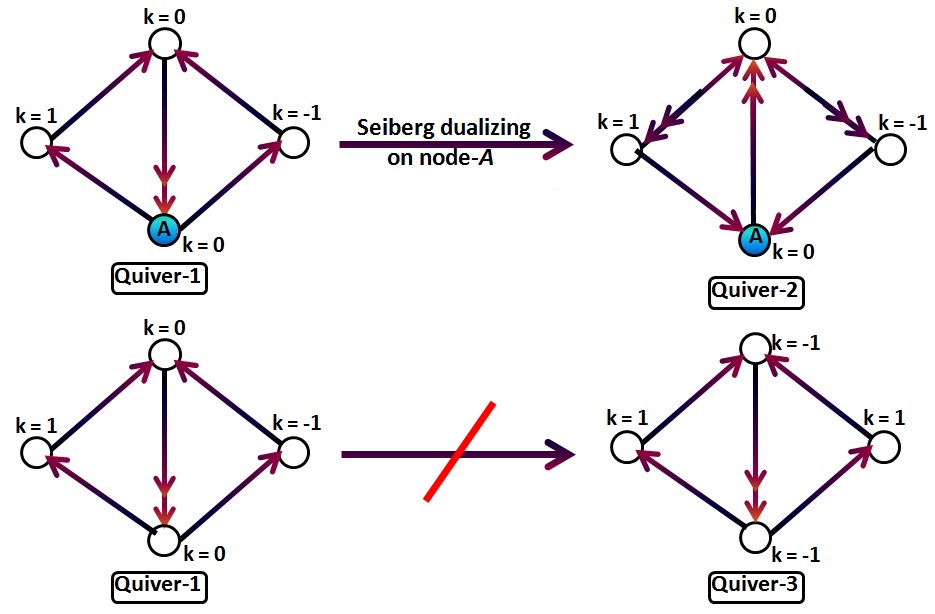}
	\caption{Quivers 1, 2 and 3 share the same toric data of $Q^{1,1,1}$. Applying the Seiberg-like transformation on node-A of quiver-1, we will get quiver-2 \cite{Amariti:2009rb}. However quiver-1 can not be transformed to quiver-3. But these quiver theories 1 and 3 have  same
superconformal indices \cite{Imamura:2011uj} .}
	\label{Q111}
\end{figure}

For (2+1)-d chiral quivers which do not have (3+1)-d parent  \cite{Franco:2008um}, the situation is more subtle. 
For example, let us consider the phases of $Q^{1,1,1}$ theory \cite{Franco:2008um} as shown in figure \ref{Q111}. The quivers 1, 2 and 3 shown in figure \ref{Q111} are all toric duals or phases of $Q^{1,1,1}$ theory.
If we apply the Seiberg-like duality transformation (figure \ref{N=1-Seiberg-Duality} $+$ rules (\ref{S-like-rules})), quiver-1 will transform to quiver-2 \cite{Amariti:2009rb}, but quiver-1 can never transform to quiver-3. In ref. \cite{Imamura:2011uj}, it was shown that the quiver-1 and quiver-3 
are toric duals having same superconformal index indicating that they could be Seiberg-like duals.

 
One of the tools for checking the Seiberg duality between two superconformal theories is to compute the Witten type index, which is also known as the superconformal index. The index for (3+1)-d superconformal field theories was given in \cite{Kinney:2005ej,Romelsberger:2007ec}, which can be used to check the AdS/CFT \cite{Maldacena:1997re} conjecture. In \cite{Bhattacharya:2008bja,Kim:2009wb}, the indices were evaluated for the ABJM theory \cite{Aharony:2008ug} on the gauge theory side and for $AdS_4 \times \frac{S^7}{\mathbb{Z}_k}$ on the gravity side. The two computations of the superconformal index agreed perfectly and it was considered as a test of the ABJM proposal. For ${\cal{N}}=2$, the details about index and a workable formula are given in \cite{Imamura:2011uj,Imamura:2011su,Hwang:2011qt,Honda:2012ik,Hwang:2012jh}. One of the feature of the superconformal index which is of relevance to this paper is that the index matches for any two Seiberg dual theories. It can be therefore used as a method to check whether two toric dual theories are Seiberg duals. We have seen in figure \ref{Q111} that the toric phases may not transform into each other under Seiberg-like transformation in (2+1)-d. So the only way out to test the Seiberg-like duality between two toric phases is to compute the superconformal index for the two theories. If the indices do not match, the theories can not be Seiberg-like duals.

In this work, we will compute the superconformal indices and check for the Seiberg-like duality between the ${\cal{N}}=2$ toric dual quiver Chern-Simons theories corresponding to complex cones over toric Fano threefolds. There are 18 toric Fano threefolds, the toric data for which are listed in \cite{Davey:2011mz}. Using the dimer tiling approach, the quiver Chern-Simons theories corresponding to 14 Fanos were obtained in \cite{Davey:2011mz}. In \cite{Dwivedi:2011zm}, the quiver theories for the remaining four Fanos ($\mathbb{P}^3$, ${\cal{B}}_1$, ${\cal{B}}_2$, ${\cal{B}}_3$) were obtained using the inverse algorithm approach and they do not admit a brane tiling description. Out of these 18 toric Fanos, only three Fano theories, namely, Fanos ${\cal{C}}_3$, ${\cal{C}}_5$ and ${\cal{B}}_2$ have phases. Complex cone over Fano ${\cal{C}}_3$ is also known as $Q^{1,1,1}/ \mathbb{Z}_2$ (sometimes denoted as $Q^{2,2,2}$) in the literature and it admits two phases \cite{Davey:2011mz} as shown in figure \ref{C3}. Superconformal indices for phases of Fano ${\cal{C}}_3$ have been evaluated in \cite{Imamura:2011uj} and they match. The phases of ${\cal{C}}_5$ have the same quiver diagram and superpotential as that of Fano ${\cal{C}}_3$ but with different Chern-Simons levels \cite{Davey:2011mz}. Besides these two Fanos, we also have Fano ${\cal{B}}_2$ whose phases \cite{Dwivedi:2011zm} are shown in figure \ref{B2}. In this paper, we will evaluate the indices for the phases of Fanos ${\cal{C}}_5$ and ${\cal{B}}_2$ to check for the Seiberg-like duality. We find that the indices for the two phases of Fano ${\cal{B}}_2$ (figure \ref{B2}(a) and figure  \ref{B2}(b)) do not match, which suggests that these two phases are not Seiberg-like dual theories.

We further study whether we could obtain toric phases by applying Seiberg-like duality transformation. In particular, we take two examples: quiver theories corresponding to phase-I of Fano ${\cal{C}}_5$
and  Fano ${\cal{B}}_3$. We find that we can not get a new toric phase for Fano ${\cal{B}}_3$ using Seiberg-like duality transformation rules.

The plan of the paper is as follows. In section \ref{sec2}, we briefly review the superconformal index for the ${\cal{N}}=2$ quiver Chern-Simons theory, giving a workable formula to compute the index, following the notations and symbols used in \cite{Imamura:2011uj}. In section \ref{sec3}, we explain the steps on how to obtain the index by taking the example of Fano ${\cal{C}}_3$ and reproducing the results of \cite{Imamura:2011uj}. We will then proceed to do the new calculations by computing the indices for phases of Fano ${\cal{C}}_5$ and Fano ${\cal{B}}_2$ and quote the results in section \ref{sec4} and section \ref{sec5} respectively. In section \ref{sec6}, we discuss the Seiberg-like duality transformation on the quivers corresponding to phase-I of Fano ${\cal{C}}_5$ and Fano ${\cal{B}}_3$ respectively, to check whether we can obtain toric phases by this approach. We will summarize and discuss in the concluding section \ref{sec7}. 
\section{Large $N$ index for ${\cal{N}}=2$ Chern-Simons}
\label{sec2}
The ${\cal{N}}=2$ superconformal index is given by \cite{Imamura:2011uj},
\sid
I(x,z_i) = \text{Tr}\left[(-1)^F x^{\Delta + j_3}{z_i}^{F_i}\right] ~,
\sidd
where the trace is taken over local gauge invariant operators. Here, $F$, $\Delta$, $j_3$ and $F_i$'s  are the fermion number, energy (or conformal dimension), projection of spin and the charge of a flavor symmetry with fugacity $z_i$'s respectively. 

A workable formula for the index can be derived from the path integral approach on $S^2 \times S^1$ using the localization methods \cite{Kim:2009wb, Imamura:2011su}. The complete index for a ${\cal{N}}=2$ superconformal toric quiver Chern-Simons theory,
with $n_G$ nodes and group $G=\prod_{A=1}^{n_G} U(N)_A$,  is given by \cite{Imamura:2011uj},
\sid
I(x,z_i) = \sum_m \int [da] \; e^{-S_{CS}} e^{i b_0(a)} x^{\epsilon_0} {z_i}^{q_{0i}} \; \text{exp}\left[\sum_{n=1}^{\infty}\frac{1}{n} f\left(e^{ina}, x^n, z_i^n\right) \right] ~,
\label{SCI}
\sidd
where `$m$' represents $n_G N$  magnetic monopole charges denoted by the set $\{m_{A,i}\}$  and `$a$' represents holonomy (Wilson line around $S^1$). The $N$ magnetic charges for each node can be arranged in a descending order (to incorporate
Weyl equivalence) and the corresponding integral over $a$ can be rewritten as:
\sid
\int da \rightarrow {\rm (const)} \left(\prod_{A=1}^{n_G} \prod_{\rho \in N_A}
\int d\rho(a)\right)  ={\rm (const)} \left(\prod_{A=1}^{n_G} \prod_{i=1}^N \int da_{A,i}\right) ~, 
\sidd 
where we pick a component of `$a$' belonging to a $U(1)$ subgroup of $U(N)_A$ by applying a weight $\rho \in N_A$. 
Here $N_A$ represents fundamental representation of $U(N)_A$ corresponding to node-A.   
The `$f$' in the argument of the exponential in eq. (\ref{SCI}) is conventionally called as letter index and it gets contributions from chiral multiplet ($f_{\text{chiral}}$) and vector multiplet ($f_{\text{vector}}$). 

For the toric quivers with bifundamental chiral multiplets directed from node-A to node-B ($\Phi_{AB}$), the explicit expressions for $f_{\text{chiral}}$ and $f_{\text{vector}}$ are as follows \cite{Imamura:2011uj}:
\begin{align}
f_{\text{chiral}}\left(e^{ia}, x, z_i\right) = & \sum_{\Phi_{AB}} \sum_{\rho \in N_A} \sum_{{\rho}^{'} \in N_B} \frac{x^{\left|\rho(m)-{\rho}^{'}(m)\right|}}{1-x^2}  \notag \\
& \times \left( e^{i[\rho(a)-{\rho}^{'}(a)]}{z_i}^{F_i(\Phi)}x^{\Delta(\Phi)} - e^{i[{\rho}^{'}(a) -\rho(a)]}{z_i}^{-F_i(\Phi)}x^{2-\Delta(\Phi)}\right)~, \notag \\
f_{\text{vector}}\left(e^{ia}, x\right) = & \sum_{A=1}^{n_G} \sum_{\rho \in N_A} \sum_{\rho^{'} \in N_A} \left(1- \delta_{\rho,{\rho}^{'}}\right) \left(-e^{i[\rho(a)-{\rho}^{'}(a)]} \; x^{\left|\rho(m)-{\rho}^{'}(m)\right|}\right) ~.
\label{letterindex} 
\end{align}
The term $S_{CS}$ is the classical contribution from Chern-Simons term, which is given by \cite{Imamura:2011uj},
\sid
S_{CS} = i \sum_{A=1}^{n_G} \sum_{\rho \in N_A} k_A \; \rho(a) \; \rho(m) ~,
\label{Sclassical}
\sidd
where $k_A$ denotes the Chern-Simons level for the node-$A$. 

All those terms in eq. (\ref{SCI}) which have subscript `0' are called zero-point contributions, where $x^{\epsilon_0}$, ${z_i}^{q_{0i}}$ and $e^{i b_0(a)}$ correspond to the zero-point energy, zero-point flavor charges and zero-point gauge charges, respectively. Their contribution is given by \cite{Imamura:2011uj}:
\begin{align}
\epsilon_0 = & \frac{1}{2} \sum_{\Phi_{AB}} \sum_{\rho \in N_A} \sum_{{\rho}^{'} \in N_B} \left|\rho(m)-{\rho}^{'}(m)\right| \left(1-\Delta(\Phi)\right) - \frac{1}{2}\sum_{A=1}^{n_G} \sum_{\rho \in N_A} \sum_{{\rho}^{'} \in N_A} \left|\rho(m)-{\rho}^{'}(m)\right| ~, \notag \\
q_{0i} = & -\frac{1}{2} \sum_{\Phi_{AB}} \sum_{\rho \in N_A} \sum_{{\rho}^{'} \in N_B} \left|\rho(m)-{\rho}^{'}(m)\right|F_i(\Phi)  ~, \notag \\
b_0(a) = & -\frac{1}{2} \sum_{\Phi_{AB}} \sum_{\rho \in N_A} \sum_{{\rho}^{'} \in N_B} \left|\rho(m)-{\rho}^{'}(m)\right| \left(\rho(a)-{\rho}^{'}(a) \right) ~.
\label{zero-point}
\end{align}

The $N$ magnetic charges $m_{A,i}$ associated with every gauge group $U(N)_A$ can be grouped into three sets: $\{m_A^{(+)}\}$ which are positive integers, $\{m_A^{(-)}\}$ which are negative integers and $\{m_A^{(0)}\}=0$ and
it has been shown that the index (\ref{SCI}) in large $N$ limit can be factorized as \cite{Imamura:2011uj}:
\sid
I = I^{(0)} I^{(+)} I^{(-)} ~, 
\sidd
where $I^{(0)}$ is independent of the magnetic charge $m$, $I^{(\pm)}$ depend only on the positive and negative parts of magnetic charges ($m^{(+)}$ and $m^{(-)}$) respectively and can be obtained by summing up the contributions of all possible $m^{(+)}$ and $m^{(-)}$, i.e.,
\sid
I^{(\pm)} = \sum_{m^{(\pm)}} I_{m^{(\pm)}}^{(\pm)} ~.
\sidd
Clearly, $I^{(+)}$ and $I^{(-)}$ will have infinite terms  to be summed. The index $I(x,z_i)$  will be a power series in 
$x$.  Truncating the series to a finite order in $x$ results in computing 
only finite number of  $I_{m^{(+)}}^{(+)}$ and $I_{m^{(-)}}^{(-)}$ in the above summation.

For convenience, we could represent the set of positive and negative charges $\{m_A^{(\pm)}\}$ corresponding to node-A 
by Young diagrams $\pm \left\{Y_A\right\}$:
\sid
\left\{m_{A}^{(\pm)}\right\} = \pm \left\{\bullet, {\tiny\yng(1)}, {\tiny\yng(2)}, {\tiny\yng(1,1)}, {\tiny\yng(2,1)}, {\tiny\yng(3,1)}, {\tiny\yng(2,2)},\ldots \right\}~.
\sidd
Note that on the right hand side, we have $\pm$ just to differentiate between $m_{A}^{(+)}$ and $m_{A}^{(-)}$.
The descending order of the positive magnetic charges $m_{A,i}^{(+)}$  will be the number of boxes 
in each row of the  Young diagram. For example, the positive magnetic charges of node-A compactly shown by 
the Young diagram $Y_A={\tiny {\yng(4,2,1)}}$ is $m_{A,1}^{(+)}=4, m_{A,2}^{(+)}=2, m_{A,3}^{(+)}=1$. 
Corresponding to this diagram for the magnetic charge, we will have integration over three holonomy variables $a_{A,1}^{(+)}, a_{A,2}^{(+)}, a_{A,3}^{(+)}$. Though the Young diagrams for every node in the quiver can be independently taken, 
we restrict to a class, called diagonal monopole operators \cite{Imamura:2011uj}, where total number of boxes of Young diagram is same 
for all nodes, that is, ($|Y_A|=|Y_B|=\ldots$). The reason is that the examples which we have considered in this paper have non-vanishing two-cycles. The non-diagonal monopole
operators correspond to the $M2$-branes wrapped on these two-cycles on the dual gravity side. These are non-BPS states and do not contribute to the index \cite{Imamura:2011uj}. So we keep only diagonal monopole operators in the index computation \cite{Imamura:2011uj}.

Suppose we take a two node quiver with positive magnetic charge as $m_{(+)} = \left(Y_A, Y_B \right) = \left({\tiny\yng(3)}, {\tiny\yng(2,1)}\right)$. We see that total number of boxes is 3 for both $Y_A$ and $Y_B$. Notice that
there will be only one holonomy variable for node-A and two holonomy variables for node-B. We will elaborate this 
approach for a four node quiver corresponding to Fano ${\cal C}_3$ in the next section.   
For completeness, we  put forth the explicit expressions for $I^{(0)}, I_{m^{(+)}}^{(+)}$ and $I_{m^{(-)}}^{(-)}$ : 
\begin{align}
I^{(0)} & = \int [d{\rho}{(a)}] e^{-S_{CS}^{(0)}} \; e^{i b_0^{(0)}} x^{\epsilon_0^{(0)}} {z_i}^{q_{0i}^{(0)}} \;  \text{exp}\left[\sum_{n=1}^{\infty}\frac{1}{n} f^{(0)}\left(e^{ina}, x^n, z_i^n\right) \right] ~, \notag \\
I_{m^{(+)}}^{(+)} & = \int [d{\rho}{(a)}] e^{-S_{CS}^{(+)}} \; e^{i b_0^{(+)}} x^{\epsilon_0^{(+)}} {z_i}^{q_{0i}^{(+)}} \text{exp}\left[\sum_{n=1}^{\infty}\frac{1}{n} f^{(+)}\left(e^{ina}, x^n, z_i^n\right) \right] ~, \notag \\
I_{m^{(-)}}^{(-)} & = \int [d{\rho}{(a)}] e^{-S_{CS}^{(-)}} \; e^{i b_0^{(-)}} x^{\epsilon_0^{(-)}} {z_i}^{q_{0i}^{(-)}} \text{exp}\left[\sum_{n=1}^{\infty}\frac{1}{n} f^{(-)}\left(e^{ina}, x^n, z_i^n\right) \right] ~.
\label{SCI0}
\end{align}
For the neutral part ($I^{(0)}$), the Chern-Simons action and zero-point contributions vanish, i.e., 
$S_{CS}^{(0)} = b_0^{(0)} = \epsilon_0^{(0)} = q_{0i}^{(0)} = 0$ and the vector and chiral contributions to letter index  $f^{(0)}$ are given below \cite{Imamura:2011uj}:
\begin{align}
f_{\text{vector}}^{(0)}\left(e^{ia}, x\right) & = -\sum_{A=1}^{n_G} \sum_{\rho \in N_A} \sum_{{\rho}^{'} \in N_A}  \left(x^{\left|\rho(m)\right|+ \left|{\rho}^{'}(m)\right|}\right)e^{i[\rho(a)-{\rho}^{'}(a)]} ~, \notag \\
f_{\text{chiral}}^{(0)}\left(e^{ia}, x, z_i\right) & = \sum_{\Phi_{AB}} \sum_{\rho \in N_A}^{(0)} \sum_{{\rho}^{'} \in N_B}^{(0)} \frac{x^{\left(\left|\rho(m)\right|+ \left|{\rho}^{'}(m)\right|\right)}}{1-x^2} \notag \\
& \times \left( e^{i[\rho(a)-{\rho}^{'}(a)]}{z_i}^{F_i(\Phi)}x^{\Delta(\Phi)} - e^{i[{\rho}^{'}(a) -\rho(a)]}{z_i}^{-F_i(\Phi)}x^{2-\Delta(\Phi)}\right) ~.
\label{SCI0-contributions}
\end{align}
Similarly for the positive and negative magnetic charges, Chern-Simons action and zero-point contributions to $I_{m^{(+)}}^{(+)}$ and $I_{m^{(-)}}^{(-)}$ are \cite{Imamura:2011uj}:
\begin{align}
 S_{CS}^{(\pm)} & = i \sum_{A=1}^{n_G} \sum_{\rho \in N_A}^{(\pm)} k_A \; \rho(a) \; \rho(m) ~, \notag \\
b_0^{(\pm)}(a) & =  - \frac{1}{2} \sum_{\Phi_{AB}} \sum_{\rho \in N_A}^{(\pm)} \sum_{{\rho}^{'} \in N_B}^{(\pm)} \left( \left|\rho(m)-{\rho}^{'}(m)\right| - \left|\rho(m)\right| - \left|{\rho}^{'}(m)\right|\right) \left(\rho(a)-{\rho}^{'}(a) \right) ~, \notag \\
\epsilon_0^{(\pm)} &= \frac{1}{2} \sum_{\Phi_{AB}} \sum_{\rho \in N_A}^{(\pm)} \sum_{{\rho}^{'} \in N_B}^{(\pm)} \left( \left|\rho(m)-{\rho}^{'}(m)\right| - \left|\rho(m)\right| - \left|{\rho}^{'}(m)\right|\right)  \left(1-\Delta(\Phi)\right) \notag \\
& - \frac{1}{2}\sum_{A=1}^{n_G} \sum_{\rho \in N_A}^{(\pm)} \sum_{{\rho}^{'} \in N_A}^{(\pm)} \left( \left|\rho(m)-{\rho}^{'}(m)\right| - \left|\rho(m)\right| - \left|{\rho}^{'}(m)\right|\right) ~, \notag \\
q_{0i}^{(\pm)} & = - \frac{1}{2} \sum_{\Phi_{AB}} \sum_{\rho \in N_A}^{(\pm)} \sum_{{\rho}^{'} \in N_B}^{(\pm)} \left( \left|\rho(m)-{\rho}^{'}(m)\right| - \left|\rho(m)\right| - \left|{\rho}^{'}(m)\right|\right) F_i(\Phi)  ~.
\label{SCS+}
\end{align}
Further the vector and chiral field contributions to letter index $f^{(\pm)}$ will be:
\begin{align}
f_{\text{vector}}^{(\pm)}\left(e^{ia}, x\right) & = \sum_{A=1}^{n_G} \sum_{\rho \in N_A}^{(\pm)} \sum_{{\rho}^{'} \in N_A}^{(\pm)} \left[ -\left(1- \delta_{\rho,\rho^{'}}\right)x^{\left|\rho(m)-{\rho}^{'}(m)\right|} + x^{\left|\rho(m)\right|+ \left|{\rho}^{'}(m)\right|} \right] e^{i[\rho(a)-{\rho}^{'}(a)]} ~, \notag \\
f_{\text{chiral}}^{(\pm)}\left(e^{ia}, x, z_i\right) & = \sum_{\Phi_{AB}} \sum_{\rho \in N_A}^{(\pm)} \sum_{{\rho}^{'} \in N_B}^{(\pm)} \frac{\left(x^{\left|\rho(m)- \rho^{'}(m)\right|} - x^{\left(\left|\rho(m)\right|+ \left|{\rho}^{'}(m)\right|\right)} \right)}{1-x^2} \notag \\
& \times \left( e^{i[\rho(a)-{\rho}^{'}(a)]}{z_i}^{F_i(\Phi)}x^{\Delta(\Phi)} - e^{i[{\rho}^{'}(a) -\rho(a)]}{z_i}^{-F_i(\Phi)}x^{2-\Delta(\Phi)}\right) ~.
\label{f+}
\end{align}
In the case of $I^{(0)}$ (\ref{SCI0}), the integral over ${\rho}{(a)}$ after substituting (\ref{SCI0-contributions}) can be written as a Gaussian integral \cite{Imamura:2011uj}. Hence the expression for $I^{(0)}$ involves the determinant of $n_G \times n_G$ matrix $M$ as follows:
\sid
I^{(0)}=\frac{1}{\prod \limits_{n=1}^{\infty }\text{Det}\left[M\left(x^n,z_i{}^n\right)\right]} ~,
\label{I0}
\sidd
where the off-diagonal and diagonal elements of $M$ are given by \cite{Imamura:2011uj},
\begin{align}
M_{AB} \; (\text{off-diagonal}) &= \frac{1}{1-x^2 }\left(\sum _{\Phi _{AB}} \left(z_i{}^{F_i\left(\Phi _{AB}\right)}x^{ \Delta \left(\Phi _{AB}\right)}\right)-\sum _{\Phi _{BA}} \left(z_i{}^{- F_i\left(\Phi _{BA}\right)}x^{2-\Delta \left(\Phi _{BA}\right)}\right)\right) ~, \notag \\
M_{AA} \; (\text{diagonal}) &= \left(-1+\frac{1}{1-x^2 }\left(\sum _{\Phi _{AA}} \left(z_i{}^{F_i} \; x^{\Delta }- z_i{}^{-F_i} \; x^{2-\Delta }\right)\right) \right) ~.
\label{MAB}
\end{align}
Here $\Phi _{AA}$ represents an adjoint field which transforms in the adjoint representation under the gauge group $U(N)_A$.

In the following sections, we will use the formula and notations given here, to obtain the indices for phases of Fanos ${\cal{C}}_3$, ${\cal{C}}_5$ and ${\cal{B}}_2$. Particularly, we will evaluate in detail, the index of phase-I of Fano ${\cal{C}}_3$ in the next section.
\section{Fano ${\cal{C}}_3$ theory}
\label{sec3}
We will  compute the superconformal index, using the formula presented in the previous section, for the two toric
phases of quiver Chern-Simons theory corresponding to Fano ${\cal{C}}_3$. The results match with the index 
presented in \cite{Imamura:2011uj}.
\begin{figure}
	\centering
		\includegraphics[width=0.75\textwidth]{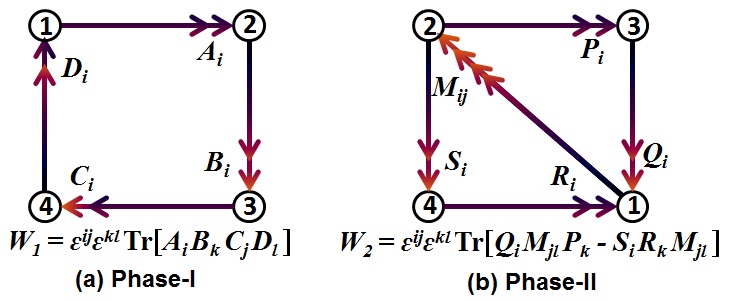}
	\caption{Quiver diagrams and superpotentials ($W_1$, $W_2$) for phase-I and phase-II of Fano ${\cal{C}}_3$.}
	\label{C3}
\end{figure}
\subsection{Index for Phase-I of Fano ${\cal{C}}_3$}
The quiver diagram and superpotential ($W_1$) for phase-I of Fano ${\cal{C}}_3$ is given in figure \ref{C3}(a). The Chern-Simons level assigned to the four nodes is $\vec{k_1} = (1,-1,-1,1)$. The $\Delta$ charge of the fields can be taken to be same as their $R$-charges. For an interaction term in the superpotential, the $R$-charge is always two. For every term in $W_1$, we see that $\Delta(A_i)+\Delta(B_i)+\Delta(C_i)+\Delta(D_i)=2$. We choose $\Delta = 1/2$ for all the fields to simplify the computation. 

We will compute the  index as a series in only one parameter $x$ by setting the other fugacities $z_i$'s as 1 and putting all flavor charges
to zero ($F_i(\Phi)=0$). These indices can be called unrefined indices. It may happen that we find the unrefined index equal for two 
different toric quivers which are not toric phases. In such a case, we have to turn on the fugacities and work out `refined index' \cite{Imamura:2011uj} and most likely the refined indices will be different. We will encounter such a situation, when we compute the unrefined index for Fano ${\cal{C}}_5$.

We change the integration variable ($\rho(a) \rightarrow e^{i \rho(a)}$) in eq. (\ref{SCI0}) which will convert the integration into a contour integration:
$$\int [d \rho(a)] \; \text{Integrand}(e^{i \rho(a)}) \longrightarrow \oint \frac{[d(e^{i \rho(a)})]}{2\pi i[e^{i \rho(a)}]} \; \text{Integrand}(e^{i \rho(a)}) ~.$$

As a warmup exercise, we evaluate the $I_{m^{(+)}}^{(+)}$ for phase-I of Fano ${\cal{C}}_3$. The theory has four gauge group factors $U(N)_1 \times U(N)_2 \times U(N)_3 \times U(N)_4$ and corresponding Chern-Simons levels are $(k_1, k_2, k_3, k_4) = (1,-1,-1,1)$. Corresponding to these gauge group factors, we will have Young diagrams $Y_1$, $Y_2$, $Y_3$ and $Y_4$ respectively. Recall that the total number of boxes in $Y_i$'s associated to the four gauge groups is same as we are also interested in
diagonal monopole operators \cite{Imamura:2011uj}. As explained with a two node  example in section \ref{sec2},  
the number of rows in each of them can be different. Let us assume that the number of rows in $Y_1$ is $P$. So each row of $Y_1$ is specified by $\rho(m)$ and $e^{i \rho(a)}$, where, $\rho(m)$ will indicate the number of boxes in any row and $e^{i \rho(a)}$ will be the corresponding integration variable. Thus, $\rho(m)$ will take values, say, $l_1$, $l_2$, ..., $l_P$, where $l_i$ denotes the number of boxes in $i$-th row of $Y_1$. Correspondingly, we  take  $e^{i \rho(a)}$ variables as $u_1$, $u_2$, ..., $u_P$, where $u_i$ is the integration variable corresponding to $l_i$. Same will be true for $Y_2$, $Y_3$ and $Y_4$. Thus, for each gauge group factor, we will assign  three labels: (Young diagram, $\rho(m)$, $e^{i \rho(a)}$).  For the Phase I of Fano ${\cal C}_3$, with four nodes, we will use the following notations for these labels:
$$ U(N)_1 : \left(Y_1, l_p, u_p\right), \; \text{$p$ runs from 1 to $P$, where $P$ is total number of rows in $Y_1$} ~.$$ 
$$ U(N)_2 : \left(Y_2, m_q, v_q\right), \; \text{$q$ runs from 1 to $Q$, where $Q$ is total number of rows in $Y_2$} ~.$$
$$ U(N)_3 : \left(Y_3, n_r, w_r\right), \; \text{$r$ runs from 1 to $R$, where $R$ is total number of rows in $Y_3$} ~.$$
$$ U(N)_4 : \left(Y_4, o_s, t_s\right), \; \text{$s$ runs from 1 to $S$, where $S$ is total number of rows in $Y_4$} ~.$$
The $I^{(+)}(x)$ will be sum of the indices for all valid Young diagrams (i.e., the total number of boxes in $Y_1, Y_2, Y_3, Y_4$ is equal):
$$I^{(+)}(x) = \sum_{Y_1,Y_2,Y_3,Y_4}I_{(Y_1,Y_2,Y_3,Y_4)}^{(+)}(x) ~.$$
The contribution of Chern-Simons action $S_{CS}^{(+)}$ (\ref{SCS+}) in $I_{(Y_1,Y_2,Y_3,Y_4)}^{(+)}$ will be:
\begin{align}
e^{-S_{CS}^{(+)}} & = \text{exp}\left[- i \sum_{A=1}^{n_G} \sum_{\rho \in N_A}^{+} k_A \rho(a) \rho(m) \right] \notag \\ 
& = \text{exp}\left[ -\left(i\sum_{\rho \in Y_1}^{+} \rho(a) \rho(m)\right) + \left(i\sum_{\rho \in Y_2}^{+} \rho(a) \rho(m)\right) \right. \notag \\
&  \hspace{1.35cm} + \left. \left(i\sum_{\rho \in Y_3}^{+} \rho(a) \rho(m)\right) -  \left(i\sum_{\rho \in Y_4}^{+} \rho(a) \rho(m)\right)\right] \notag \\
& =  \text{exp} \left[-\left(i\sum_{p = 1}^{P} \rho(a) l_p\right) + \left(i\sum_{q = 1}^{Q} \rho(a) m_q\right) + \left(i\sum_{r = 1}^{R} \rho(a) n_r\right) - \left(i\sum_{s = 1}^{S} \rho(a) o_s\right) \right] \notag \\
& = \left(\prod _{p=1}^P u_p^{-l_p}\right)\left(\prod _{q=1}^Q v_q^{m_q}\right)\left(\prod _{r=1}^R w_r^{n_r}\right)\left(\prod _{s=1}^S t_s^{-o_s}\right) ~,
\end{align}
where $\sum_{\rho \in N_A}^{+}$ denotes summing over $\rho$ for the positive magnetic charges.

The contribution to $b_0^{(+)}$ (\ref{SCS+}) from the bifundamental field $A_1$ (from node-1 to node-2 as shown in figure \ref{C3}(a)) will be:
\begin{align}
b_0^{(+)} (A_1) & = - \frac{1}{2} \sum_{\rho \in Y_1}^{(+)} \sum_{{\rho}^{'} \in Y_2}^{(+)} \left( \left|\rho(m)-{\rho}^{'}(m)\right| - \left|\rho(m)\right| - \left|{\rho}^{'}(m)\right|\right) \left(\rho(a)-{\rho}^{'}(a) \right) \notag  \\ 
& =  - \frac{1}{2} \sum_{p=1}^{P} \sum_{q=1}^{Q} \left( \left|l_p-m_q\right| - \left|l_p\right| - \left|m_q\right|\right) \left(\rho(a)-{\rho}^{'}(a) \right) \notag \\
\implies e^{i b_0^{(+)}}(A_1) & = \prod _{p=1}^P \prod _{q=1}^Q  {\left(u_p^{-1} v_q \right)}^{\frac{1}{2} \left(\left|l_p-m_q\right|-\left|l_p\right|-\left|m_q\right|\right)} ~.
\end{align}
Taking the contributions from all bifundamental fields, we obtain:
\begin{align}
e^{i b_0^{(+)}} & = \left( \prod _{p=1}^P \prod _{q=1}^Q  {\left(u_p^{-1} v_q \right)}^{\left(\left|l_p-m_q\right|-\left|l_p\right|-\left|m_q\right|\right)} \right) \times \left( \prod _{q=1}^Q \prod _{r=1}^R  {\left(v_q^{-1} w_r \right)}^{ \left(\left|m_q-n_r\right|-\left|m_q\right|-\left|n_r\right|\right)} \right) \notag \\     
& \times \left( \prod _{r=1}^R \prod _{s=1}^S  {\left(w_r^{-1} t_s \right)}^{\left(\left|n_r-o_s\right|-\left|n_r\right|-\left|o_s\right|\right)} \right) \times \left( \prod _{s=1}^S \prod _{p=1}^P  {\left(t_s^{-1} u_p \right)}^{\left(\left|o_s-l_p\right|-\left|o_s\right|-\left|l_p\right|\right)} \right) ~.
\end{align} 

Similarly, the expression for $\epsilon_0^{(+)}$ (\ref{SCS+}) will be given as:
\begin{align}
\epsilon_0^{(+)} & = \frac{1}{2} \left(\sum _{p=1}^P \sum _{q=1}^Q \left(\left|l_p-m_q\right|-\left|l_p\right|-\left|m_q\right|\right)\right) + \frac{1}{2} \left(\sum _{q=1}^Q \sum _{r=1}^R \left(\left|m_q-n_r\right|-\left|m_q\right|-\left|n_r\right|\right)\right)  \notag \\ 
& + \frac{1}{2} \left(\sum _{r=1}^R \sum _{s=1}^S \left(\left|n_r-o_s\right|-\left|n_r\right|-\left|o_s\right|\right)\right) + \frac{1}{2} \left(\sum _{s=1}^S \sum _{p=1}^P \left(\left|o_s-l_p\right|-\left|o_s\right|-\left|l_p\right|\right)\right) \notag \\
& - \frac{1}{2} \left(\sum _{i=1}^P \sum _{j=1}^P \left(\left|l_i-l_j\right|-\left|l_i\right|-\left|l_j\right|\right)\right) - \frac{1}{2} \left(\sum _{i=1}^Q \sum _{j=1}^Q \left(\left|m_i-m_j\right|-\left|m_i\right|-\left|m_j\right|\right)\right) \notag \\
& - \frac{1}{2} \left(\sum _{i=1}^R \sum _{j=1}^R \left(\left|n_i-n_j\right|-\left|n_i\right|-\left|n_j\right|\right)\right) - \frac{1}{2} \left(\sum _{i=1}^S \sum _{j=1}^S \left(\left|o_i-o_j\right|-\left|o_i\right|-\left|o_j\right|\right)\right) ~.
\end{align}  

Substituting the vector ($f_{\text{vector}}^{(+)}$) and chiral ($f_{\text{chiral}}^{(+)}$) parts from eq. (\ref{f+}) in the exponential part of $I_{m^{(+)}}^{(+)}$ given in eq. (\ref{SCI0}), we can write the complete integral for $I_{m^{(+)}}^{(+)}$ as:
\begin{align}
I_{(Y_1, Y_2, Y_3, Y_4)}^{(+)} & = \oint \prod _{p=1}^P\left(\frac{du_p}{2 \pi i u_p}\right)\prod _{q=1}^Q\left(\frac{dv_q}{2 \pi i v_q}\right)\prod _{r=1}^R\left(\frac{dw_r}{2 \pi i w_r}\right)\prod _{s=1}^S\left(\frac{dt_s}{2 \pi i t_s}\right)  \notag \\
& \times \left( e^{-S_{CS}^{(+)}} e^{i b_0^{(+)}} x^{\epsilon_0^{(+)}} F_{\text{vector}}^{(+)} F_{\text{chiral}}^{(+)} \right) ~.
\label{complete-integral}
\end{align}
where $F_{\text{vector}}^{(+)}$ and $F_{\text{chiral}}^{(+)}$ are the Plethystic exponential of $f_{\text{vector}}^{(+)}$ and $f_{\text{chiral}}^{(+)}$ which can be written as:
\begin{align}
F_{\text{vector}}^{(+)}  = & \left(\prod _{i=1}^P \prod _{j=1}^P \frac{\left(1-(1-\delta_{i,j})x^{\left|l_i-l_j\right|} u_iu_j{}^{-1}\right)}{\left(1-x^{\left(\left|l_i\right|+\left|l_j\right|\right)} u_iu_j^{-1}\right)}\right) \left(\prod _{i=1}^Q \prod _{j=1}^Q \frac{\left(1-(1-\delta_{i,j})x^{\left|m_i-m_j\right|} v_iv_j{}^{-1}\right)}{\left(1-x^{\left(\left|m_i\right|+\left|m_j\right|\right)} v_iv_j^{-1}\right)}\right) \notag \\
& \left(\prod _{i=1}^R \prod _{j=1}^R \frac{\left(1-(1-\delta_{i,j})x^{\left|n_i-n_j\right|} w_iw_j{}^{-1}\right)}{\left(1-x^{\left(\left|n_i\right|+\left|n_j\right|\right)} w_iw_j^{-1}\right)}\right) \left(\prod _{i=1}^S \prod _{j=1}^S \frac{\left(1-(1-\delta_{i,j})x^{\left|o_i-o_j\right|} t_it_j{}^{-1}\right)}{\left(1-x^{\left(\left|o_i\right|+\left|o_j\right|\right)} t_it_j^{-1}\right)}\right) ~, \notag \\
F_{\text{chiral}}^{(+)}  = &
\left(\prod _{d=0}^{\infty} \left(\prod _{p=1}^P \prod _{q=1}^Q \frac{\left(1-x^{\left(2d+\left|l_p-m_q\right|+\frac{3}{2}\right)} u_p^{-1}v_q\right)\left(1-x^{\left(2d+\left|l_p\right|+\left|m_q\right|+\frac{1}{2}\right)} u_pv_q^{-1}\right)}{\left(1-x^{\left(2d+\left|l_p-m_q\right|+\frac{1}{2}\right)} u_pv_q^{-1}\right)\left(1-x^{2d+\left|l_p\right|+\left|m_q\right|+\frac{3}{2}} u_p^{-1}v_q\right)}\right)^2\right) \notag \\
& \left(\prod _{d=0}^{\infty} \left(\prod _{q=1}^Q \prod _{r=1}^R \frac{\left(1-x^{\left(2d+\left|m_q-n_r\right|+\frac{3}{2}\right)} v_q^{-1}w_r\right)\left(1-x^{\left(2d+\left|m_q\right|+\left|n_r\right|+\frac{1}{2}\right)} v_qw_r^{-1}\right)}{\left(1-x^{\left(2d+\left|m_q-n_r\right|+\frac{1}{2}\right)} v_qw_r^{-1}\right)\left(1-x^{2d+\left|m_q\right|+\left|n_r\right|+\frac{3}{2}} v_q^{-1}w_r\right)}\right)^2\right) \notag \\
& \left(\prod _{d=0}^{\infty} \left(\prod _{r=1}^R \prod _{s=1}^S \frac{\left(1-x^{\left(2d+\left|n_r-o_s\right|+\frac{3}{2}\right)} w_r^{-1}t_s\right)\left(1-x^{\left(2d+\left|n_r\right|+\left|o_s\right|+\frac{1}{2}\right)} w_rt_s^{-1}\right)}{\left(1-x^{\left(2d+\left|n_r-o_s\right|+\frac{1}{2}\right)} w_rt_s^{-1}\right)\left(1-x^{2d+\left|n_r\right|+\left|o_s\right|+\frac{3}{2}} w_r^{-1}t_s\right)}\right)^2\right) \notag \\
& \left(\prod _{d=0}^{\infty} \left(\prod _{s=1}^S \prod _{p=1}^P \frac{\left(1-x^{\left(2d+\left|o_s-l_p\right|+\frac{3}{2}\right)} t_s^{-1}u_p\right)\left(1-x^{\left(2d+\left|o_s\right|+\left|l_p\right|+\frac{1}{2}\right)} t_su_p^{-1}\right)}{\left(1-x^{\left(2d+\left|o_s-l_p\right|+\frac{1}{2}\right)} t_su_p^{-1}\right)\left(1-x^{2d+\left|o_s\right|+\left|l_p\right|+\frac{3}{2}} t_s^{-1}u_p\right)}\right)^2\right) \notag~.
\end{align} 
Though these expressions have infinite products, we will truncate to certain power of $x$, say $x^5$, which will result in computing finite number of terms in product. 

Having obtained all the factors, we can perform the integration, for a particular choice of $Y_1, Y_2, Y_3, Y_4$, by obtaining the residues at the origin for all the integration variables. We can have the following cases:
\begin{itemize}
\item \underline{Number of boxes in $Y_1$, $Y_2$, $Y_3$, $Y_4$ $= 0$}
\newline Here, $P=Q=R=S=0$. So the only possibility is $Y_1 = Y_2 = Y_3 = Y_4 = \bullet$ and we have $I_{(\bullet,\bullet,\bullet,\bullet)}^{(+)} = 1$.
\item \underline{Number of boxes in $Y_1$, $Y_2$, $Y_3$, $Y_4$ $= 1$}
\newline Since we have one box, only one row is possible. Hence, $P=Q=R=S=1$ and $Y_1 = Y_2 = Y_3 = Y_4 = \tiny\yng(1)$. Up to order $x^5$, we obtain,
\sid
I_{\left(\tiny\yng(1),\; \tiny\yng(1),\; \tiny\yng(1),\; \tiny\yng(1)\right)}^{(+)} = 8 x^2 ~.
\sidd
\item \underline{Number of boxes in $Y_1$, $Y_2$, $Y_3$, $Y_4$ $= 2$}
\newline As the total number of boxes are two, so each of $Y_1, Y_2, Y_3, Y_4$ can have either one or two rows. So we can have total $2^4$ possibilities. However, $I_{(Y_1, Y_2, Y_3, Y_4)}^{(+)}$ will be same up to permutations of $Y_1, Y_2, Y_3, Y_4$. Thus we are only required to evaluate $I_{\left(\tiny\yng(2),\; \tiny\yng(2),\; \tiny\yng(2),\; \tiny\yng(2)\right)}^{(+)}$, $I_{\left(\tiny\yng(1,1),\; \tiny\yng(2),\; \tiny\yng(2),\; \tiny\yng(2)\right)}^{(+)}$, $I_{\left(\tiny\yng(1,1),\; \tiny\yng(1,1),\; \tiny\yng(2),\; \tiny\yng(2)\right)}^{(+)}$, $I_{\left(\tiny\yng(1,1),\; \tiny\yng(1,1),\; \tiny\yng(1,1),\; \tiny\yng(2)\right)}^{(+)}$ and $I_{\left(\tiny\yng(1,1),\; \tiny\yng(1,1),\; \tiny\yng(1,1),\; \tiny\yng(1,1)\right)}^{(+)}$. The contributions of these up to order $x^5$ are given below: 
\sid
I_{\left(\tiny\yng(2),\; \tiny\yng(2),\; \tiny\yng(2),\; \tiny\yng(2)\right)}^{(+)} = 16x^4 ~,
\sidd
$$I_{\left(\tiny\yng(1,1),\; \tiny\yng(1,1),\; \tiny\yng(2),\; \tiny\yng(2)\right)}^{(+)} = I_{\left(\tiny\yng(1,1),\; \tiny\yng(2),\; \tiny\yng(2),\; \tiny\yng(2)\right)}^{(+)} = I_{\left(\tiny\yng(1,1),\; \tiny\yng(1,1),\; \tiny\yng(1,1),\; \tiny\yng(2)\right)}^{(+)} = I_{\left(\tiny\yng(1,1),\; \tiny\yng(1,1),\; \tiny\yng(1,1),\; \tiny\yng(1,1)\right)}^{(+)} = 0~.$$
\end{itemize}
The Young tableau with higher number of rows and boxes will not contribute up to order $x^5$.
If we set the flavor fugacities to 1, then $I^{(-)}(x)$ will be same as $I^{(+)}(x)$.

Let us now calculate $I^{(0)}$ (\ref{I0}). Using the simplifications given in this section, the elements of matrix $M$ (\ref{MAB}), which is a $4 \times 4$ matrix, are given by:  
\begin{align}
M_{AB} & =\frac{1}{1-x^2 }\left(\sum _{\Phi _{AB}} x^{ \Delta \left(\Phi _{AB}\right)}-\sum _{\Phi _{BA}} x^{2-\Delta \left(\Phi _{BA}\right)}\right) = \frac{1}{1-x^2 }\left(\sum _{\Phi _{AB}} x^{1/2}-\sum _{\Phi _{BA}} x^{3/2}\right) ~, \notag \\
M_{AA} & =\left(-1+\frac{1}{1-x^2 }\left(\sum _{\Phi _{AA}} \left(x^{\Delta }-x^{2-\Delta }\right)\right) \right) = -1 ~.
\end{align}
Hence the matrix $M$ will be given by,
\sid
M(x) =\left(
\begin{array}{cccc}
 -1 & \frac{2x^{1/2}}{1-x^2} & 0 & -\frac{2x^{3/2}}{1-x^2} \\
 -\frac{2x^{3/2}}{1-x^2} & -1 & \frac{2x^{1/2}}{1-x^2} & 0 \\
 0 & -\frac{2x^{3/2}}{1-x^2} & -1 & \frac{2x^{1/2}}{1-x^2} \\
 \frac{2x^{1/2}}{1-x^2} & 0 & -\frac{2x^{3/2}}{1-x^2} & -1
\end{array}
\right) ~.
\sidd
The determinant of this matrix $\left|M\left(x^n\right)\right| = 1$. Thus, from (\ref{I0}), we get $I^{(0)}(x) = 1$. 
The main purpose of the detailed steps is to provide a clear understanding of the index computation.  
This will help the readers to work out index for other quivers.
In the following subsection, we will present the unrefined index up to order $x^5$ for another toric phase of Fano ${\cal C}_3$.   
\subsection{Index for Phase-II of Fano ${\cal{C}}_3$}
The quiver and superpotential ($W_2$) for phase-II of Fano ${\cal{C}}_3$ is given in figure \ref{C3}(b). The Chern-Simons levels are $\vec{k_2} = (-1,-1,1,1)$. The $\Delta$ charge of the fields can be taken as $\Delta = 2/3$. Using the systematic approach given for the phase-I, one can evaluate the indices for phase-II of Fano ${\cal{C}}_3$. The non-zero contributions up to order $x^5$ are given below:
\sid
I_{\left(\tiny\yng(1),\; \tiny\yng(1),\; \tiny\yng(1),\; \tiny\yng(1)\right)}^{(+)} = 8 x^2, \quad I_{\left(\tiny\yng(2),\; \tiny\yng(2),\; \tiny\yng(2),\; \tiny\yng(2)\right)}^{(+)} = 16x^4, \quad I^{(0)} = 1 ~,
\sidd
which is same as corresponding indices for phase-I of Fano ${\cal{C}}_3$. It is appropriate to mention that the refined indices, with the flavor fugacities ($z_i$) turned on, for the two phases also agree \cite{Imamura:2011uj}.  
\section{Indices for the two phases of Fano ${\cal{C}}_5$}
\label{sec4}
The quiver diagram and the superpotential for phase-I and phase-II of Fano ${\cal{C}}_5$ is exactly the same as that of Fano ${\cal{C}}_3$ as shown in figure \ref{C3}(a) and figure \ref{C3}(b). However the Chern-Simons levels are different \cite{Davey:2011mz} which are $\vec{k_1}=(1,-2,1,0)$ and $\vec{k_2}=(0,0,1,-1)$ for phase-I and phase-II of Fano ${\cal{C}}_5$ respectively. The non-zero contributions to the unrefined index up to order $x^5$ are: 
\sid
\text{\underline{For phase-I:}} \quad I_{\left(\tiny\yng(1),\; \tiny\yng(1),\; \tiny\yng(1),\; \tiny\yng(1)\right)}^{(+)} = 8 x^2, \quad I_{\left(\tiny\yng(2),\; \tiny\yng(2),\; \tiny\yng(2),\; \tiny\yng(2)\right)}^{(+)} = 16x^4, \quad I^{(0)} = 1 ~.
\sidd
\sid
\text{\underline{For phase-II:}} \quad I_{\left(\tiny\yng(1),\; \tiny\yng(1),\; \tiny\yng(1),\; \tiny\yng(1)\right)}^{(+)} = 8 x^2, \quad I_{\left(\tiny\yng(2),\; \tiny\yng(2),\; \tiny\yng(2),\; \tiny\yng(2)\right)}^{(+)} = 16x^4, \quad I^{(0)} = 1 ~.
\sidd
Thus, the unrefined index for the two phases agree. Further they seem to accidentally match with the corresponding indices for phases of Fano ${\cal{C}}_3$. But we have checked by turning on fugacities, that the refined indices for Fano ${\cal{C}}_5$ is not the same as that of Fano ${\cal{C}}_3$.
\section{Indices for the two phases of Fano ${\cal{B}}_2$}
\label{sec5}
\begin{figure}
	\centering
		\includegraphics[width=0.80\textwidth]{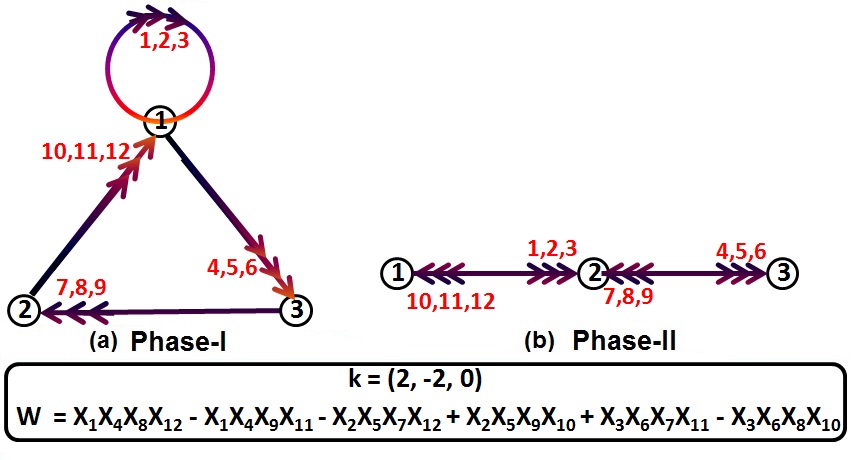}
	\caption{The two phases of Fano $\Bb$ theory. Both the phases have same Chern-Simons level and superpotential with different quiver diagrams.}
	\label{B2}
\end{figure}
This theory was obtained in \cite{Dwivedi:2011zm} and it does not admit a dimer tiling description. The theory has 2 toric phases whose quivers are shown in figure \ref{B2}(a) and figure \ref{B2}(b) respectively. Both the phases have Chern-Simons level $\vec{k} = (2,-2,0)$ and superpotential,
\sid
 W = X_1 X_4X_8X_{12} -X_1 X_4X_9X_{11} - X_2X_5X_7X_{12} + X_2 X_5 X_9 X_{10} + X_3 X_6X_7X_{11} - X_3 X_6X_8X_{10} ~. \notag 
\sidd
The explicit evaluation for $I_{(Y_1, Y_2, Y_3)}^{(+)}$ and $I^{(0)}$ for some of the ($Y_1, Y_2, Y_3$), keeping terms up to $x^2$ have been performed for the two phases and results are given below.
\vspace{0.4cm}
\newline
\textbf{Index for Phase-I of Fano ${\cal{B}}_2$}
\begin{align}
I_{\tiny\yng(1),\; \tiny\yng(1),\; \tiny\yng(1)}^{(+)} = & 138+\frac{36}{x}+\frac{81}{\sqrt{x}}+399 \sqrt{x}+369 x+363 x^{3/2}+891 x^2 ~,\notag \\
I_{\tiny\yng(2),\; \tiny\yng(2),\; \tiny\yng(2)}^{(+)} = & 825+\frac{225}{x^2}+\frac{375}{x^{3/2}}+\frac{558}{x}+\frac{1413}{\sqrt{x}}+1365 \sqrt{x}+3030 x+2094 x^{3/2}+3375 x^2 ~, \notag \\
I^{(0)} = & 1+3 \sqrt{x}+12 x+63 x^{3/2}+246 x^2 ~.
\label{phase-I-B2}
\end{align}
\newline
\textbf{Index for Phase-II of Fano ${\cal{B}}_2$}
\begin{align}
I_{\tiny\yng(1),\; \tiny\yng(1),\; \tiny\yng(1)}^{(+)} = & -2+\frac{1}{x^2}+\frac{3}{x}-x + 4x^2 ~, \notag \\
I_{\tiny\yng(2),\; \tiny\yng(2),\; \tiny\yng(2)}^{(+)} = & 3+\frac{1}{x^4}+\frac{3}{x^3}-\frac{4}{x^2}-\frac{2}{x}-24 x + 28x^2 ~, \notag \\
I^{(0)} = & 1 + 18 x + 306 x^2 ~.
\label{phase-II-B2}
\end{align}
Comparing the results of eqs. (\ref{phase-I-B2}) and (\ref{phase-II-B2}), we see that the indices for the two phases do not match. So we can conclude that phase-I and phase-II of Fano ${\cal{B}}_2$ are not Seiberg-like dual theories.  
\section{Seiberg-like duality transformation and toric phases} \label{sec6}
In this section, we will study the Seiberg-like duality transformation (\ref{S-like-rules}) on the quiver Chern-Simons theories corresponding to Phase-I of Fano ${\cal{C}}_5$ and Fano ${\cal{B}}_3$ and check if we can find the toric phases for these quivers. The Seiberg-like duals of phase-I of Fano ${\cal{C}}_5$ has already been discussed in \cite{Amariti:2009rb}. In the following subsection, we will briefly review it. 
\subsection{Seiberg-like dual theory of phase-I of Fano ${\cal{C}}_5$}
\begin{figure}
	\centering
		\includegraphics[width=0.85\textwidth]{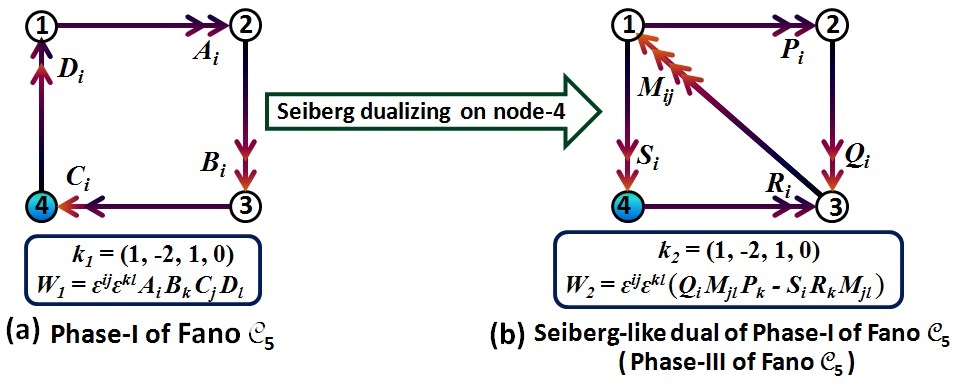}
	\caption{Seiberg-like duality on node-4 of phase-I of Fano ${\cal{C}}_5$. The dual theory obtained is toric dual to phase-I, which we call as phase-III of Fano ${\cal{C}}_5$.}
	\label{C5-Seiberg}
\end{figure}

The quiver diagram for phase-I of Fano ${\cal{C}}_5$  is shown in figure \ref{C5-Seiberg}(a). The Chern-Simons level is $\vec{k_1} = (1,-2,1,0)$ and superpotential is,
\sid
W_1 = \epsilon^{ij}\epsilon^{kl}\left(A_iB_kC_jD_l\right) ~.
\sidd
This phase has a (3+1)-d parent \cite{Feng:2001bn}. The Seiberg duality on this phase was done in \cite{Feng:2001bn} to get the toric phase which is the (3+1)-d parent of quiver diagram shown in figure \ref{C5-Seiberg}(b). In \cite{Amariti:2009rb}, the Chern-Simons levels were assigned appropriately on these two Seibeg-like dual theories in accordance with rules (\ref{S-like-rules}) so that they become toric duals. Applying Seiberg-like duality on node-4 of phase-I of Fano ${\cal{C}}_5$, we get a theory as shown in figure \ref{C5-Seiberg}(b) which has Chern-Simons level $\vec{k_2} = (1,-2,1,0)$ and superpotential,
\sid
W_2 = \epsilon^{ij}\epsilon^{kl}\left(Q_i M_{jl} P_k - S_i R_k M_{jl} \right) ~.
\sidd  
This theory is a toric quiver Chern-Simons theory. Moreover, for this choice of Chern-Simons level, the toric data obtained turns out to be related by a $GL(4,\mathbb{Z})$ transformation, with the toric data of phase-I of Fano ${\cal{C}}_5$ \cite{Amariti:2009rb}. So, these two theories are also toric duals. This approach has led to obtaining a toric phase \cite{Amariti:2009rb} of Fano ${\cal{C}}_5$ which is different from phase-I and phase-II, which we call as phase-III of Fano ${\cal{C}}_5$ as shown in figure \ref{C5-Seiberg}. This phase-III is a member of the more general family discussed in \cite{Amariti:2009rb}.
\subsection{Seiberg-like dual theory of Fano ${\cal{B}}_3$}
\begin{figure}
	\centering
		\includegraphics[width=0.85\textwidth]{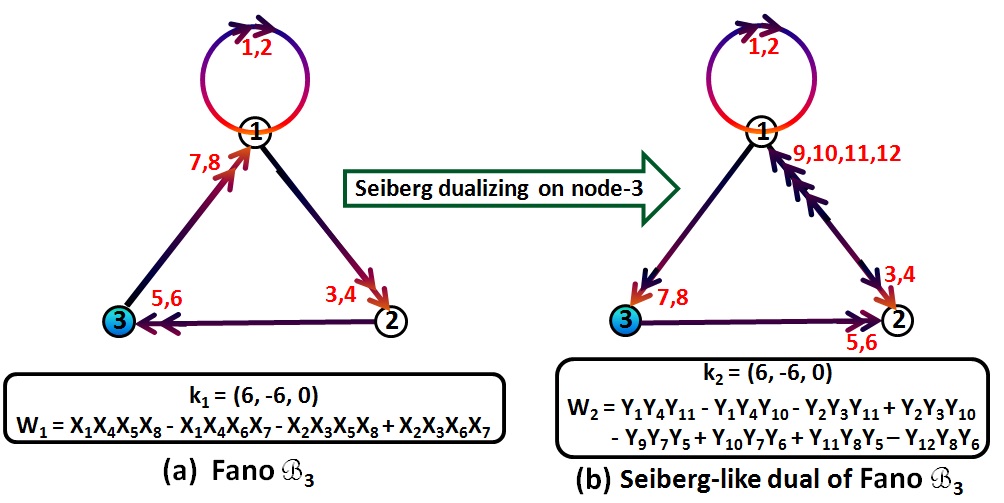}
	\caption{Seiberg-like duality on node-3 of Fano $\Bc$ theory. We obtain the Seiberg-like dual theory whose superpotential ($W_2$) is not toric.}
	\label{B3}
\end{figure}

The quiver diagram for Fano ${\cal{B}}_3$ theory is given in figure \ref{B3}(a). This theory has Chern-Simons level $\vec{k_1} = (6,-6,0)$ and the superpotential as,
\sid 
W_1 = X_1 X_4X_5X_8 - X_1 X_4X_6X_7 -X_2X_3X_5X_8+X_2X_3X_6X_7 ~.
\sidd
In the quiver diagram of Fano ${\cal{B}}_3$, we see that node-3 has level 0 and also has two incoming and two outgoing arrows as shown in figure \ref{B3}(a). So, we perform the Seiberg-like duality on node-3. Using the steps illustrated in figure \ref{N=1-Seiberg-Duality} and rules given in eq. (\ref{S-like-rules}), we obtain the Seiberg-like dual of Fano ${\cal{B}}_3$, the quiver diagram of which is shown in figure \ref{B3}(b). This dual theory has same Chern-Simons level $\vec{k_2} = (6,-6,0)$ and we obtain the new superpotential ($W_2$) following ref. \cite{Feng:2001bn} to be,
\begin{align} 
W_2 & = Y_1 Y_4 Y_{11} - Y_1 Y_4 Y_{10} - Y_2 Y_3 Y_{11} + Y_2 Y_3 Y_{10} \notag \\ 
& - Y_9 Y_7 Y_{5} + Y_{10} Y_7 Y_6  + Y_{11} Y_8 Y_5 - Y_{12} Y_8 Y_6 ~, 
\label{W-dualB3}
\end{align}
where, $Y_1$, $Y_2$ are the adjoints on node-1 and remaining fields $Y_i (3 \leq i \leq 12)$ are the bifundamental fields as shown in the figure \ref{B3}(b). From eq. (\ref{W-dualB3}), we see that fields $Y_9,Y_{10},Y_{11},Y_{12}$ do not appear exactly twice in $W_2$. Hence the superpotential $W_2$ is not toric and hence, the Seiberg-like dual theory of Fano ${\cal{B}}_3$ is not a toric quiver Chern-Simons theory.  

From the examples of Fano ${\cal{B}}_3$ and phase-I of Fano ${\cal{C}}_5$ worked out in this section, we cannot expect Seiberg-like duals in (2+1)-d to be toric duals, in general.
\section{Conclusion}
\label{sec7}
In this work, we have reviewed some of the issues relating toric duality and Seiberg-like duality in (2+1)-dimensions for the ${\cal{N}}=2$ quiver Chern-Simons theories corresponding to complex cones over Fano threefolds. In particular, we have focused on Fanos ${\cal{C}}_3$, ${\cal{C}}_5$ and ${\cal{B}}_2$, which are the only Fanos in the literature, known to admit phases \cite{Davey:2011mz,Dwivedi:2011zm}. We try to find out whether these phases are also Seiberg-like duals. The usual Seiberg-like duality transformation in (2+1)-d can not be trusted for chiral quivers. For example, the phases shown by quiver-1 and quiver-3 in figure \ref{Q111} do not transform into each other by rules (\ref{S-like-rules}). However, the superconformal indices at large $N$ for these theories, up to second order in $x$ \cite{Imamura:2011uj}, seem to match. 

We work out the indices for phases of Fanos ${\cal{C}}_3$, ${\cal{C}}_5$, ${\cal{B}}_2$ to see if they agree. Our computation of large $N$ indices for phases of Fano ${\cal{C}}_3$ is consistent with the calculations done in \cite{Imamura:2011uj}. In section \ref{sec4} and section \ref{sec5}, we have done new calculations by evaluating the superconformal indices for the phases of Fano ${\cal{C}}_5$ and Fano ${\cal{B}}_2$ respectively. We find that the indices for phases of Fano ${\cal{B}}_2$ do not match, which suggests that these two toric dual theories are not Seiberg-like duals.

We have also performed the Seiberg-like transformation on the quivers corresponding to phase-I of Fano ${\cal{C}}_5$ and Fano ${\cal{B}}_3$ as shown in figure \ref{C5-Seiberg} and figure \ref{B3} respectively, to check if this method can give toric phases. In case of Fano ${\cal{B}}_3$, the new Seiberg-like dual theory obtained has a superpotential given in eq. (\ref{W-dualB3}), which does not satisfy the toric condition. It means that the dual theory obtained is not a toric quiver gauge theory. This suggests that Seiberg-like duality may or may not be toric duality. It would be an interesting exercise to check if the Seiberg-like duality can give new phases for other Fano threefold theories, which were studied in \cite{Davey:2011mz}.

In this paper, we have worked out unrefined large $N$ superconformal indices for toric phases. Except for phase-II of
Fano $\Bb$, all the quivers we considered are chiral quivers. 
We hope to pursue $AdS_4/CFT_3$ checks for the chiral quivers
in future. 

It is pertinent to mention some of the issues about chiral quivers. The free energy for chiral quivers 
scale as $N^2$ instead of the $N^{3/2}$ \cite{Jafferis:2011zi, Amariti:2011uw, Gang:2011jj, Kim:2012vza} which is required to correctly reproduce the volume of the dual 7-dimensional Sasaki-Einstein manifold ($SE_7$) . However in \cite{Amariti:2011jp}, a symmetrization technique was proposed to address this free energy scaling problem. This technique was used for phase-I of Fano ${\cal{C}}_3$ with Chern-Simons level $(k,-k,k,-k)$ and also for Fano $\Bd$ with Chern-Simons level $(k,-2k,k)$ and the correct volumes of $Q^{2,2,2} /\BZ_k$ and $M^{1,1,1} /\BZ_k$ were reproduced \cite{Amariti:2011uw, Gang:2011jj}.
Further for chiral theories, there is a discrepancy between the gauge theory superconformal index and the index on the corresponding gravity side at higher orders \cite{Kim:2010vwa,Cheon:2011th}. We hope to figure out whether there is
procedure  analogous to symmetrization procedure which will help in resolving this discrepancy. For Fano ${\cal{C}}_3$
and ${\cal{C}}_5$, our index computation upto order $x^2$ matched with the gravity index \cite{Eager:2012hx, Eager:2013mua}. We
are presently working on refined index beyond order $x^2$ for chiral quivers and comparing with the corresponding
gravity index.

\acknowledgments
The authors would like to thank Abhijit Gadde for useful discussions and technical details on the topic. We would also like to acknowledge the email correspondence with Masazumi Honda and Yoshinori Honma regarding some calculation aspects. S.D. would like to thank the organizers of National Strings Meeting (NSM-2013) at IIT Kharagpur, where this work was presented. We are thankful to Richard Eager for bringing to our notice about their work on gravity superconformal indices. 

\begin{thebibliography}{49}
\expandafter\ifx\csname natexlab\endcsname\relax\def\natexlab#1{#1}\fi
\expandafter\ifx\csname bibnamefont\endcsname\relax
  \def\bibnamefont#1{#1}\fi
\expandafter\ifx\csname bibfnamefont\endcsname\relax
  \def\bibfnamefont#1{#1}\fi
\expandafter\ifx\csname citenamefont\endcsname\relax
  \def\citenamefont#1{#1}\fi
\expandafter\ifx\csname url\endcsname\relax
  \def\url#1{\texttt{#1}}\fi
\expandafter\ifx\csname urlprefix\endcsname\relax\def\urlprefix{URL }\fi
\providecommand{\bibinfo}[2]{#2}
\providecommand{\eprint}[2][]{\url{#2}}

\bibitem[{\citenamefont{Franco et~al.}(2006{\natexlab{a}})\citenamefont{Franco,
  Hanany, Martelli, Sparks, Vegh et~al.}}]{Franco:2005sm}
\bibinfo{author}{\bibfnamefont{S.}~\bibnamefont{Franco}},
  \bibinfo{author}{\bibfnamefont{A.}~\bibnamefont{Hanany}},
  \bibinfo{author}{\bibfnamefont{D.}~\bibnamefont{Martelli}},
  \bibinfo{author}{\bibfnamefont{J.}~\bibnamefont{Sparks}},
  \bibinfo{author}{\bibfnamefont{D.}~\bibnamefont{Vegh}}, \bibnamefont{et~al.},
  \bibinfo{journal}{JHEP} \textbf{\bibinfo{volume}{0601}}, \bibinfo{pages}{128}
  (\bibinfo{year}{2006}{\natexlab{a}}), \eprint{hep-th/0505211}.

\bibitem[{\citenamefont{Feng et~al.}(2001{\natexlab{a}})\citenamefont{Feng,
  Hanany, and He}}]{Feng:2000mi}
\bibinfo{author}{\bibfnamefont{B.}~\bibnamefont{Feng}},
  \bibinfo{author}{\bibfnamefont{A.}~\bibnamefont{Hanany}}, \bibnamefont{and}
  \bibinfo{author}{\bibfnamefont{Y.-H.} \bibnamefont{He}},
  \bibinfo{journal}{Nucl.Phys.} \textbf{\bibinfo{volume}{B595}},
  \bibinfo{pages}{165} (\bibinfo{year}{2001}{\natexlab{a}}),
  \eprint{hep-th/0003085}.

\bibitem[{\citenamefont{Ueda and Yamazaki}(2008)}]{Ueda:2008hx}
\bibinfo{author}{\bibfnamefont{K.}~\bibnamefont{Ueda}} \bibnamefont{and}
  \bibinfo{author}{\bibfnamefont{M.}~\bibnamefont{Yamazaki}},
  \bibinfo{journal}{JHEP} \textbf{\bibinfo{volume}{0812}}, \bibinfo{pages}{045}
  (\bibinfo{year}{2008}), \eprint{0808.3768}.

\bibitem[{\citenamefont{Hanany et~al.}(2009)\citenamefont{Hanany, Vegh, and
  Zaffaroni}}]{Hanany:2008fj}
\bibinfo{author}{\bibfnamefont{A.}~\bibnamefont{Hanany}},
  \bibinfo{author}{\bibfnamefont{D.}~\bibnamefont{Vegh}}, \bibnamefont{and}
  \bibinfo{author}{\bibfnamefont{A.}~\bibnamefont{Zaffaroni}},
  \bibinfo{journal}{JHEP} \textbf{\bibinfo{volume}{0903}}, \bibinfo{pages}{012}
  (\bibinfo{year}{2009}), \eprint{0809.1440}.

\bibitem[{\citenamefont{Franco et~al.}(2008)\citenamefont{Franco, Hanany, Park,
  and Rodriguez-Gomez}}]{Franco:2008um}
\bibinfo{author}{\bibfnamefont{S.}~\bibnamefont{Franco}},
  \bibinfo{author}{\bibfnamefont{A.}~\bibnamefont{Hanany}},
  \bibinfo{author}{\bibfnamefont{J.}~\bibnamefont{Park}}, \bibnamefont{and}
  \bibinfo{author}{\bibfnamefont{D.}~\bibnamefont{Rodriguez-Gomez}},
  \bibinfo{journal}{JHEP} \textbf{\bibinfo{volume}{0812}}, \bibinfo{pages}{110}
  (\bibinfo{year}{2008}), \eprint{0809.3237}.

\bibitem[{\citenamefont{Hanany and He}(2008)}]{Hanany:2008gx}
\bibinfo{author}{\bibfnamefont{A.}~\bibnamefont{Hanany}} \bibnamefont{and}
  \bibinfo{author}{\bibfnamefont{Y.-H.} \bibnamefont{He}}
  (\bibinfo{year}{2008}), \eprint{0811.4044}.

\bibitem[{\citenamefont{Davey et~al.}(2009{\natexlab{a}})\citenamefont{Davey,
  Hanany, Mekareeya, and Torri}}]{Davey:2009et}
\bibinfo{author}{\bibfnamefont{J.}~\bibnamefont{Davey}},
  \bibinfo{author}{\bibfnamefont{A.}~\bibnamefont{Hanany}},
  \bibinfo{author}{\bibfnamefont{N.}~\bibnamefont{Mekareeya}},
  \bibnamefont{and} \bibinfo{author}{\bibfnamefont{G.}~\bibnamefont{Torri}}
  (\bibinfo{year}{2009}{\natexlab{a}}), \eprint{0910.4962}.

\bibitem[{\citenamefont{Feng et~al.}(2001{\natexlab{b}})\citenamefont{Feng,
  Hanany, and He}}]{Feng:2001xr}
\bibinfo{author}{\bibfnamefont{B.}~\bibnamefont{Feng}},
  \bibinfo{author}{\bibfnamefont{A.}~\bibnamefont{Hanany}}, \bibnamefont{and}
  \bibinfo{author}{\bibfnamefont{Y.-H.} \bibnamefont{He}},
  \bibinfo{journal}{JHEP} \textbf{\bibinfo{volume}{0108}}, \bibinfo{pages}{040}
  (\bibinfo{year}{2001}{\natexlab{b}}), \eprint{hep-th/0104259}.

\bibitem[{\citenamefont{Feng et~al.}(2002)\citenamefont{Feng, Franco, Hanany,
  and He}}]{Feng:2002zw}
\bibinfo{author}{\bibfnamefont{B.}~\bibnamefont{Feng}},
  \bibinfo{author}{\bibfnamefont{S.}~\bibnamefont{Franco}},
  \bibinfo{author}{\bibfnamefont{A.}~\bibnamefont{Hanany}}, \bibnamefont{and}
  \bibinfo{author}{\bibfnamefont{Y.-H.} \bibnamefont{He}},
  \bibinfo{journal}{JHEP} \textbf{\bibinfo{volume}{0212}}, \bibinfo{pages}{076}
  (\bibinfo{year}{2002}), \eprint{hep-th/0205144}.

\bibitem[{\citenamefont{Davey et~al.}(2009{\natexlab{b}})\citenamefont{Davey,
  Hanany, Mekareeya, and Torri}}]{Davey:2009sr}
\bibinfo{author}{\bibfnamefont{J.}~\bibnamefont{Davey}},
  \bibinfo{author}{\bibfnamefont{A.}~\bibnamefont{Hanany}},
  \bibinfo{author}{\bibfnamefont{N.}~\bibnamefont{Mekareeya}},
  \bibnamefont{and} \bibinfo{author}{\bibfnamefont{G.}~\bibnamefont{Torri}},
  \bibinfo{journal}{JHEP} \textbf{\bibinfo{volume}{0906}}, \bibinfo{pages}{025}
  (\bibinfo{year}{2009}{\natexlab{b}}), \eprint{0903.3234}.

\bibitem[{\citenamefont{Davey et~al.}(2009{\natexlab{c}})\citenamefont{Davey,
  Hanany, Mekareeya, and Torri}}]{Davey:2009qx}
\bibinfo{author}{\bibfnamefont{J.}~\bibnamefont{Davey}},
  \bibinfo{author}{\bibfnamefont{A.}~\bibnamefont{Hanany}},
  \bibinfo{author}{\bibfnamefont{N.}~\bibnamefont{Mekareeya}},
  \bibnamefont{and} \bibinfo{author}{\bibfnamefont{G.}~\bibnamefont{Torri}},
  \bibinfo{journal}{JHEP} \textbf{\bibinfo{volume}{0911}}, \bibinfo{pages}{028}
  (\bibinfo{year}{2009}{\natexlab{c}}), \eprint{0908.4033}.

\bibitem[{\citenamefont{Hanany and Kennaway}(2005)}]{Hanany:2005ve}
\bibinfo{author}{\bibfnamefont{A.}~\bibnamefont{Hanany}} \bibnamefont{and}
  \bibinfo{author}{\bibfnamefont{K.~D.} \bibnamefont{Kennaway}}
  (\bibinfo{year}{2005}), \eprint{hep-th/0503149}.

\bibitem[{\citenamefont{Franco et~al.}(2006{\natexlab{b}})\citenamefont{Franco,
  Hanany, Kennaway, Vegh, and Wecht}}]{Franco:2005rj}
\bibinfo{author}{\bibfnamefont{S.}~\bibnamefont{Franco}},
  \bibinfo{author}{\bibfnamefont{A.}~\bibnamefont{Hanany}},
  \bibinfo{author}{\bibfnamefont{K.~D.} \bibnamefont{Kennaway}},
  \bibinfo{author}{\bibfnamefont{D.}~\bibnamefont{Vegh}}, \bibnamefont{and}
  \bibinfo{author}{\bibfnamefont{B.}~\bibnamefont{Wecht}},
  \bibinfo{journal}{JHEP} \textbf{\bibinfo{volume}{0601}}, \bibinfo{pages}{096}
  (\bibinfo{year}{2006}{\natexlab{b}}), \eprint{hep-th/0504110}.

\bibitem[{\citenamefont{Hanany and Vegh}(2007)}]{Hanany:2005ss}
\bibinfo{author}{\bibfnamefont{A.}~\bibnamefont{Hanany}} \bibnamefont{and}
  \bibinfo{author}{\bibfnamefont{D.}~\bibnamefont{Vegh}},
  \bibinfo{journal}{JHEP} \textbf{\bibinfo{volume}{0710}}, \bibinfo{pages}{029}
  (\bibinfo{year}{2007}), \eprint{hep-th/0511063}.

\bibitem[{\citenamefont{Kennaway}(2007)}]{Kennaway:2007tq}
\bibinfo{author}{\bibfnamefont{K.~D.} \bibnamefont{Kennaway}},
  \bibinfo{journal}{Int.J.Mod.Phys.} \textbf{\bibinfo{volume}{A22}},
  \bibinfo{pages}{2977} (\bibinfo{year}{2007}), \eprint{0706.1660}.

\bibitem[{\citenamefont{Agarwal et~al.}(2008)\citenamefont{Agarwal, Ramadevi,
  and Sarkar}}]{Agarwal:2008yb}
\bibinfo{author}{\bibfnamefont{P.}~\bibnamefont{Agarwal}},
  \bibinfo{author}{\bibfnamefont{P.}~\bibnamefont{Ramadevi}}, \bibnamefont{and}
  \bibinfo{author}{\bibfnamefont{T.}~\bibnamefont{Sarkar}},
  \bibinfo{journal}{JHEP} \textbf{\bibinfo{volume}{0806}}, \bibinfo{pages}{054}
  (\bibinfo{year}{2008}), \eprint{0804.1902}.

\bibitem[{\citenamefont{Hanany and Zaffaroni}(2008)}]{Hanany:2008cd}
\bibinfo{author}{\bibfnamefont{A.}~\bibnamefont{Hanany}} \bibnamefont{and}
  \bibinfo{author}{\bibfnamefont{A.}~\bibnamefont{Zaffaroni}},
  \bibinfo{journal}{JHEP} \textbf{\bibinfo{volume}{0810}}, \bibinfo{pages}{111}
  (\bibinfo{year}{2008}), \eprint{0808.1244}.

\bibitem[{\citenamefont{Dwivedi and Ramadevi}(2011)}]{Dwivedi:2011zm}
\bibinfo{author}{\bibfnamefont{S.}~\bibnamefont{Dwivedi}} \bibnamefont{and}
  \bibinfo{author}{\bibfnamefont{P.}~\bibnamefont{Ramadevi}},
  \bibinfo{journal}{JHEP} \textbf{\bibinfo{volume}{1111}}, \bibinfo{pages}{111}
  (\bibinfo{year}{2011}), \eprint{1108.2387}.

\bibitem[{\citenamefont{Seiberg}(1995)}]{Seiberg:1994pq}
\bibinfo{author}{\bibfnamefont{N.}~\bibnamefont{Seiberg}},
  \bibinfo{journal}{Nucl.Phys.} \textbf{\bibinfo{volume}{B435}},
  \bibinfo{pages}{129} (\bibinfo{year}{1995}), \eprint{hep-th/9411149}.

\bibitem[{\citenamefont{Berenstein and Douglas}(2002)}]{Berenstein:2002fi}
\bibinfo{author}{\bibfnamefont{D.}~\bibnamefont{Berenstein}} \bibnamefont{and}
  \bibinfo{author}{\bibfnamefont{M.~R.} \bibnamefont{Douglas}}
  (\bibinfo{year}{2002}), \eprint{hep-th/0207027}.

\bibitem[{\citenamefont{Feng et~al.}(2001{\natexlab{c}})\citenamefont{Feng,
  Hanany, He, and Uranga}}]{Feng:2001bn}
\bibinfo{author}{\bibfnamefont{B.}~\bibnamefont{Feng}},
  \bibinfo{author}{\bibfnamefont{A.}~\bibnamefont{Hanany}},
  \bibinfo{author}{\bibfnamefont{Y.-H.} \bibnamefont{He}}, \bibnamefont{and}
  \bibinfo{author}{\bibfnamefont{A.~M.} \bibnamefont{Uranga}},
  \bibinfo{journal}{JHEP} \textbf{\bibinfo{volume}{0112}}, \bibinfo{pages}{035}
  (\bibinfo{year}{2001}{\natexlab{c}}), \eprint{hep-th/0109063}.

\bibitem[{\citenamefont{Beasley and Plesser}(2001)}]{Beasley:2001zp}
\bibinfo{author}{\bibfnamefont{C.~E.} \bibnamefont{Beasley}} \bibnamefont{and}
  \bibinfo{author}{\bibfnamefont{M.~R.} \bibnamefont{Plesser}},
  \bibinfo{journal}{JHEP} \textbf{\bibinfo{volume}{0112}}, \bibinfo{pages}{001}
  (\bibinfo{year}{2001}), \eprint{hep-th/0109053}.

\bibitem[{\citenamefont{Robles-Llana}(2004)}]{RoblesLlana:2004nq}
\bibinfo{author}{\bibfnamefont{D.}~\bibnamefont{Robles-Llana}}
  (\bibinfo{year}{2004}), \eprint{hep-th/0411059}.

\bibitem[{\citenamefont{Benini et~al.}(2011)\citenamefont{Benini, Closset, and
  Cremonesi}}]{Benini:2011mf}
\bibinfo{author}{\bibfnamefont{F.}~\bibnamefont{Benini}},
  \bibinfo{author}{\bibfnamefont{C.}~\bibnamefont{Closset}}, \bibnamefont{and}
  \bibinfo{author}{\bibfnamefont{S.}~\bibnamefont{Cremonesi}},
  \bibinfo{journal}{JHEP} \textbf{\bibinfo{volume}{1110}}, \bibinfo{pages}{075}
  (\bibinfo{year}{2011}), \eprint{1108.5373}.

\bibitem[{\citenamefont{Closset}(2012)}]{Closset:2012eq}
\bibinfo{author}{\bibfnamefont{C.}~\bibnamefont{Closset}},
  \bibinfo{journal}{JHEP} \textbf{\bibinfo{volume}{1203}}, \bibinfo{pages}{056}
  (\bibinfo{year}{2012}), \eprint{1201.2432}.

\bibitem[{\citenamefont{Agarwal et~al.}(2012)\citenamefont{Agarwal, Amariti,
  and Siani}}]{Agarwal:2012wd}
\bibinfo{author}{\bibfnamefont{P.}~\bibnamefont{Agarwal}},
  \bibinfo{author}{\bibfnamefont{A.}~\bibnamefont{Amariti}}, \bibnamefont{and}
  \bibinfo{author}{\bibfnamefont{M.}~\bibnamefont{Siani}},
  \bibinfo{journal}{JHEP} \textbf{\bibinfo{volume}{1210}}, \bibinfo{pages}{178}
  (\bibinfo{year}{2012}), \eprint{1205.6798}.

\bibitem[{\citenamefont{Park and Park}(2013)}]{Park:2013wta}
\bibinfo{author}{\bibfnamefont{J.}~\bibnamefont{Park}} \bibnamefont{and}
  \bibinfo{author}{\bibfnamefont{K.-J.} \bibnamefont{Park}}
  (\bibinfo{year}{2013}), \eprint{1305.6280}.

\bibitem[{\citenamefont{Amariti et~al.}(2012)\citenamefont{Amariti, Klare, and
  Siani}}]{Amariti:2011uw}
\bibinfo{author}{\bibfnamefont{A.}~\bibnamefont{Amariti}},
  \bibinfo{author}{\bibfnamefont{C.}~\bibnamefont{Klare}}, \bibnamefont{and}
  \bibinfo{author}{\bibfnamefont{M.}~\bibnamefont{Siani}},
  \bibinfo{journal}{JHEP} \textbf{\bibinfo{volume}{1210}}, \bibinfo{pages}{019}
  (\bibinfo{year}{2012}), \eprint{1111.1723}.

\bibitem[{\citenamefont{Amariti et~al.}(2010)\citenamefont{Amariti, Forcella,
  Girardello, and Mariotti}}]{Amariti:2009rb}
\bibinfo{author}{\bibfnamefont{A.}~\bibnamefont{Amariti}},
  \bibinfo{author}{\bibfnamefont{D.}~\bibnamefont{Forcella}},
  \bibinfo{author}{\bibfnamefont{L.}~\bibnamefont{Girardello}},
  \bibnamefont{and} \bibinfo{author}{\bibfnamefont{A.}~\bibnamefont{Mariotti}},
  \bibinfo{journal}{JHEP} \textbf{\bibinfo{volume}{1005}}, \bibinfo{pages}{025}
  (\bibinfo{year}{2010}), \eprint{0903.3222}.

\bibitem[{\citenamefont{Imamura et~al.}(2011)\citenamefont{Imamura, Yokoyama,
  and Yokoyama}}]{Imamura:2011uj}
\bibinfo{author}{\bibfnamefont{Y.}~\bibnamefont{Imamura}},
  \bibinfo{author}{\bibfnamefont{D.}~\bibnamefont{Yokoyama}}, \bibnamefont{and}
  \bibinfo{author}{\bibfnamefont{S.}~\bibnamefont{Yokoyama}},
  \bibinfo{journal}{JHEP} \textbf{\bibinfo{volume}{1108}}, \bibinfo{pages}{011}
  (\bibinfo{year}{2011}), \eprint{1102.0621}.

\bibitem[{\citenamefont{Kinney et~al.}(2007)\citenamefont{Kinney, Maldacena,
  Minwalla, and Raju}}]{Kinney:2005ej}
\bibinfo{author}{\bibfnamefont{J.}~\bibnamefont{Kinney}},
  \bibinfo{author}{\bibfnamefont{J.~M.} \bibnamefont{Maldacena}},
  \bibinfo{author}{\bibfnamefont{S.}~\bibnamefont{Minwalla}}, \bibnamefont{and}
  \bibinfo{author}{\bibfnamefont{S.}~\bibnamefont{Raju}},
  \bibinfo{journal}{Commun.Math.Phys.} \textbf{\bibinfo{volume}{275}},
  \bibinfo{pages}{209} (\bibinfo{year}{2007}), \eprint{hep-th/0510251}.

\bibitem[{\citenamefont{Romelsberger}(2007)}]{Romelsberger:2007ec}
\bibinfo{author}{\bibfnamefont{C.}~\bibnamefont{Romelsberger}}
  (\bibinfo{year}{2007}), \eprint{0707.3702}.

\bibitem[{\citenamefont{Maldacena}(1998)}]{Maldacena:1997re}
\bibinfo{author}{\bibfnamefont{J.~M.} \bibnamefont{Maldacena}},
  \bibinfo{journal}{Adv.Theor.Math.Phys.} \textbf{\bibinfo{volume}{2}},
  \bibinfo{pages}{231} (\bibinfo{year}{1998}), \eprint{hep-th/9711200}.

\bibitem[{\citenamefont{Bhattacharya and
  Minwalla}(2009)}]{Bhattacharya:2008bja}
\bibinfo{author}{\bibfnamefont{J.}~\bibnamefont{Bhattacharya}}
  \bibnamefont{and} \bibinfo{author}{\bibfnamefont{S.}~\bibnamefont{Minwalla}},
  \bibinfo{journal}{JHEP} \textbf{\bibinfo{volume}{0901}}, \bibinfo{pages}{014}
  (\bibinfo{year}{2009}), \eprint{0806.3251}.

\bibitem[{\citenamefont{Kim}(2009)}]{Kim:2009wb}
\bibinfo{author}{\bibfnamefont{S.}~\bibnamefont{Kim}},
  \bibinfo{journal}{Nucl.Phys.} \textbf{\bibinfo{volume}{B821}},
  \bibinfo{pages}{241} (\bibinfo{year}{2009}), \eprint{0903.4172}.

\bibitem[{\citenamefont{Aharony et~al.}(2008)\citenamefont{Aharony, Bergman,
  Jafferis, and Maldacena}}]{Aharony:2008ug}
\bibinfo{author}{\bibfnamefont{O.}~\bibnamefont{Aharony}},
  \bibinfo{author}{\bibfnamefont{O.}~\bibnamefont{Bergman}},
  \bibinfo{author}{\bibfnamefont{D.~L.} \bibnamefont{Jafferis}},
  \bibnamefont{and}
  \bibinfo{author}{\bibfnamefont{J.}~\bibnamefont{Maldacena}},
  \bibinfo{journal}{JHEP} \textbf{\bibinfo{volume}{0810}}, \bibinfo{pages}{091}
  (\bibinfo{year}{2008}), \eprint{0806.1218}.

\bibitem[{\citenamefont{Imamura and Yokoyama}(2011)}]{Imamura:2011su}
\bibinfo{author}{\bibfnamefont{Y.}~\bibnamefont{Imamura}} \bibnamefont{and}
  \bibinfo{author}{\bibfnamefont{S.}~\bibnamefont{Yokoyama}},
  \bibinfo{journal}{JHEP} \textbf{\bibinfo{volume}{1104}}, \bibinfo{pages}{007}
  (\bibinfo{year}{2011}), \eprint{1101.0557}.

\bibitem[{\citenamefont{Hwang et~al.}(2011)\citenamefont{Hwang, Kim, Park, and
  Park}}]{Hwang:2011qt}
\bibinfo{author}{\bibfnamefont{C.}~\bibnamefont{Hwang}},
  \bibinfo{author}{\bibfnamefont{H.}~\bibnamefont{Kim}},
  \bibinfo{author}{\bibfnamefont{K.-J.} \bibnamefont{Park}}, \bibnamefont{and}
  \bibinfo{author}{\bibfnamefont{J.}~\bibnamefont{Park}},
  \bibinfo{journal}{JHEP} \textbf{\bibinfo{volume}{1109}}, \bibinfo{pages}{037}
  (\bibinfo{year}{2011}), \eprint{1107.4942}.

\bibitem[{\citenamefont{Honda and Honma}(2013)}]{Honda:2012ik}
\bibinfo{author}{\bibfnamefont{M.}~\bibnamefont{Honda}} \bibnamefont{and}
  \bibinfo{author}{\bibfnamefont{Y.}~\bibnamefont{Honma}},
  \bibinfo{journal}{JHEP} \textbf{\bibinfo{volume}{1301}}, \bibinfo{pages}{159}
  (\bibinfo{year}{2013}), \eprint{1210.1371}.

\bibitem[{\citenamefont{Hwang et~al.}(2012)\citenamefont{Hwang, Kim, and
  Park}}]{Hwang:2012jh}
\bibinfo{author}{\bibfnamefont{C.}~\bibnamefont{Hwang}},
  \bibinfo{author}{\bibfnamefont{H.-C.} \bibnamefont{Kim}}, \bibnamefont{and}
  \bibinfo{author}{\bibfnamefont{J.}~\bibnamefont{Park}}
  (\bibinfo{year}{2012}), \eprint{1211.6023}.

\bibitem[{\citenamefont{Davey et~al.}(2011)\citenamefont{Davey, Hanany,
  Mekareeya, and Torri}}]{Davey:2011mz}
\bibinfo{author}{\bibfnamefont{J.}~\bibnamefont{Davey}},
  \bibinfo{author}{\bibfnamefont{A.}~\bibnamefont{Hanany}},
  \bibinfo{author}{\bibfnamefont{N.}~\bibnamefont{Mekareeya}},
  \bibnamefont{and} \bibinfo{author}{\bibfnamefont{G.}~\bibnamefont{Torri}},
  \bibinfo{journal}{J.Phys.} \textbf{\bibinfo{volume}{A44}},
  \bibinfo{pages}{405401} (\bibinfo{year}{2011}), \eprint{1103.0553}.

\bibitem[{\citenamefont{Jafferis et~al.}(2011)\citenamefont{Jafferis, Klebanov,
  Pufu, and Safdi}}]{Jafferis:2011zi}
\bibinfo{author}{\bibfnamefont{D.~L.} \bibnamefont{Jafferis}},
  \bibinfo{author}{\bibfnamefont{I.~R.} \bibnamefont{Klebanov}},
  \bibinfo{author}{\bibfnamefont{S.~S.} \bibnamefont{Pufu}}, \bibnamefont{and}
  \bibinfo{author}{\bibfnamefont{B.~R.} \bibnamefont{Safdi}},
  \bibinfo{journal}{JHEP} \textbf{\bibinfo{volume}{1106}}, \bibinfo{pages}{102}
  (\bibinfo{year}{2011}), \eprint{1103.1181}.

\bibitem[{\citenamefont{Gang et~al.}(2012)\citenamefont{Gang, Hwang, Kim, and
  Park}}]{Gang:2011jj}
\bibinfo{author}{\bibfnamefont{D.}~\bibnamefont{Gang}},
  \bibinfo{author}{\bibfnamefont{C.}~\bibnamefont{Hwang}},
  \bibinfo{author}{\bibfnamefont{S.}~\bibnamefont{Kim}}, \bibnamefont{and}
  \bibinfo{author}{\bibfnamefont{J.}~\bibnamefont{Park}},
  \bibinfo{journal}{JHEP} \textbf{\bibinfo{volume}{1202}}, \bibinfo{pages}{079}
  (\bibinfo{year}{2012}), \eprint{1111.4529}.

\bibitem[{\citenamefont{Kim and Kim}(2012)}]{Kim:2012vza}
\bibinfo{author}{\bibfnamefont{H.}~\bibnamefont{Kim}} \bibnamefont{and}
  \bibinfo{author}{\bibfnamefont{N.}~\bibnamefont{Kim}},
  \bibinfo{journal}{JHEP} \textbf{\bibinfo{volume}{1205}}, \bibinfo{pages}{152}
  (\bibinfo{year}{2012}), \eprint{1202.6637}.

\bibitem[{\citenamefont{Amariti and Siani}(2012)}]{Amariti:2011jp}
\bibinfo{author}{\bibfnamefont{A.}~\bibnamefont{Amariti}} \bibnamefont{and}
  \bibinfo{author}{\bibfnamefont{M.}~\bibnamefont{Siani}},
  \bibinfo{journal}{JHEP} \textbf{\bibinfo{volume}{1206}}, \bibinfo{pages}{171}
  (\bibinfo{year}{2012}), \eprint{1109.4152}.

\bibitem[{\citenamefont{Kim and Park}(2010)}]{Kim:2010vwa}
\bibinfo{author}{\bibfnamefont{S.}~\bibnamefont{Kim}} \bibnamefont{and}
  \bibinfo{author}{\bibfnamefont{J.}~\bibnamefont{Park}},
  \bibinfo{journal}{JHEP} \textbf{\bibinfo{volume}{1008}}, \bibinfo{pages}{069}
  (\bibinfo{year}{2010}), \eprint{1003.4343}.

\bibitem[{\citenamefont{Cheon et~al.}(2011)\citenamefont{Cheon, Gang, Kim, and
  Park}}]{Cheon:2011th}
\bibinfo{author}{\bibfnamefont{S.}~\bibnamefont{Cheon}},
  \bibinfo{author}{\bibfnamefont{D.}~\bibnamefont{Gang}},
  \bibinfo{author}{\bibfnamefont{S.}~\bibnamefont{Kim}}, \bibnamefont{and}
  \bibinfo{author}{\bibfnamefont{J.}~\bibnamefont{Park}},
  \bibinfo{journal}{JHEP} \textbf{\bibinfo{volume}{1105}}, \bibinfo{pages}{027}
  (\bibinfo{year}{2011}), \eprint{1102.4273}.

\bibitem[{\citenamefont{Eager et~al.}(2012)\citenamefont{Eager, Schmude, and
  Tachikawa}}]{Eager:2012hx}
\bibinfo{author}{\bibfnamefont{R.}~\bibnamefont{Eager}},
  \bibinfo{author}{\bibfnamefont{J.}~\bibnamefont{Schmude}}, \bibnamefont{and}
  \bibinfo{author}{\bibfnamefont{Y.}~\bibnamefont{Tachikawa}}
  (\bibinfo{year}{2012}), \eprint{1207.0573}.

\bibitem[{\citenamefont{Eager and Schmude}(2013)}]{Eager:2013mua}
\bibinfo{author}{\bibfnamefont{R.}~\bibnamefont{Eager}} \bibnamefont{and}
  \bibinfo{author}{\bibfnamefont{J.}~\bibnamefont{Schmude}}
  (\bibinfo{year}{2013}), \eprint{1305.3547}.

\end{thebibliography}

\end{document}